%\magnification=\magstep1
\hsize=31pc 
\vsize=49pc 
\lineskip=0pt 
\parskip=0pt plus 1pt 
\hfuzz=1pt   
\vfuzz=2pt 
\pretolerance=2500 
\tolerance=5000 
\vbadness=5000 
\hbadness=5000 
\widowpenalty=500 
\clubpenalty=200 
\brokenpenalty=500 
\predisplaypenalty=200 
\voffset=-1pc 
\nopagenumbers      
\catcode`@=11 
\newif\ifams 
\amsfalse %\amstrue si ¤ suivant utilise
% 
%%%%%%%%%%%%%%%%%%%%%%%%%%%%%%%%%%%%%%%%%%%%%%%%%%%%%%%%%%%%% 
%                                                           % 
%  The following section may be commented out and           % 
%  \ifams set to either \amstrue to use the AMS fonts       % 
%  or \amsfalse if they are not available                   % 
%                                                           % 
%%%%%%%%%%%%%%%%%%%%%%%%%%%%%%%%%%%%%%%%%%%%%%%%%%%%%%%%%%%%% 
% 
%\def\Yesreply{Y } 
%\def\Noreply{N } 
%\def\yesreply{y } 
%\def\noreply{n } 
%\newif\ifnotyorn 
%\message{Do you want to use AMSfonts, msam and msbm? Y or N: }% 
%\loop 
%\read-1 to \reply 
%\ifx\reply\yesreply\global\amstrue\notyornfalse 
%\else\ifx\reply\Yesreply\global\amstrue\notyornfalse 
%\else\ifx\reply\noreply\global\amsfalse\notyornfalse 
%\else\ifx\reply\Noreply\global\amsfalse\notyornfalse 
%\else\notyorntrue 
%\message{Please type y or Y  (Yes) or n or N (No)}\fi\fi\fi\fi 
%\ifnotyorn\repeat 
%%%%%%%%%%%%%%%%%%%%%%%%%%%%%%%%%%%%%%%%%%%%%%%%%%%%%%%%%%%% 
% 
\newfam\bdifam 
\newfam\bsyfam 
\newfam\bssfam 
\newfam\msafam 
\newfam\msbfam 
\newif\ifxxpt    
\newif\ifxviipt  
\newif\ifxivpt   
\newif\ifxiipt   
\newif\ifxipt    
\newif\ifxpt     
\newif\ifixpt    
\newif\ifviiipt  
\newif\ifviipt   
\newif\ifvipt    
\newif\ifvpt     
% 
% Headings in 20pt, 17pt or 14pt 
% 
\def\headsize#1#2{\def\headb@seline{#2}% 
                \ifnum#1=20\def\HEAD{twenty}% 
                           \def\smHEAD{twelve}% 
                           \def\vsHEAD{nine}% 
                           \ifxxpt\else\xdef\f@ntsize{\HEAD}% 
                           \def\m@g{4}\def\s@ze{20.74}% 
                           \loadheadfonts\xxpttrue\fi 
                           \ifxiipt\else\xdef\f@ntsize{\smHEAD}% 
                           \def\m@g{1}\def\s@ze{12}% 
                           \loadxiiptfonts\xiipttrue\fi 
                           \ifixpt\else\xdef\f@ntsize{\vsHEAD}% 
                           \def\s@ze{9}% 
                           \loadsmallfonts\ixpttrue\fi 
                      \else 
                \ifnum#1=17\def\HEAD{seventeen}% 
                           \def\smHEAD{eleven}% 
                           \def\vsHEAD{eight}% 
                           \ifxviipt\else\xdef\f@ntsize{\HEAD}% 
                           \def\m@g{3}\def\s@ze{17.28}% 
                           \loadheadfonts\xviipttrue\fi 
                           \ifxipt\else\xdef\f@ntsize{\smHEAD}% 
                           \loadxiptfonts\xipttrue\fi 
                           \ifviiipt\else\xdef\f@ntsize{\vsHEAD}% 
                           \def\s@ze{8}% 
                           \loadsmallfonts\viiipttrue\fi 
                      \else\def\HEAD{fourteen}% 
                           \def\smHEAD{ten}% 
                           \def\vsHEAD{seven}% 
                           \ifxivpt\else\xdef\f@ntsize{\HEAD}% 
                           \def\m@g{2}\def\s@ze{14.4}% 
                           \loadheadfonts\xivpttrue\fi 
                           \ifxpt\else\xdef\f@ntsize{\smHEAD}% 
                           \def\s@ze{10}% 
                           \loadxptfonts\xpttrue\fi 
                           \ifviipt\else\xdef\f@ntsize{\vsHEAD}% 
                           \def\s@ze{7}% 
                           \loadviiptfonts\viipttrue\fi 
                \ifnum#1=14\else 
                \message{Header size should be 20, 17 or 14 point 
                              will now default to 14pt}\fi 
                \fi\fi\headfonts} 
% 
% Text in 12pt, 11pt or 10pt  
% 
\def\textsize#1#2{\def\textb@seline{#2}% 
                 \ifnum#1=12\def\TEXT{twelve}% 
                           \def\smTEXT{eight}% 
                           \def\vsTEXT{six}% 
                           \ifxiipt\else\xdef\f@ntsize{\TEXT}% 
                           \def\m@g{1}\def\s@ze{12}% 
                           \loadxiiptfonts\xiipttrue\fi 
                           \ifviiipt\else\xdef\f@ntsize{\smTEXT}% 
                           \def\s@ze{8}% 
                           \loadsmallfonts\viiipttrue\fi 
                           \ifvipt\else\xdef\f@ntsize{\vsTEXT}% 
                           \def\s@ze{6}% 
                           \loadviptfonts\vipttrue\fi 
                      \else 
                \ifnum#1=11\def\TEXT{eleven}% 
                           \def\smTEXT{seven}% 
                           \def\vsTEXT{five}% 
                           \ifxipt\else\xdef\f@ntsize{\TEXT}% 
                           \def\s@ze{11}% 
                           \loadxiptfonts\xipttrue\fi 
                           \ifviipt\else\xdef\f@ntsize{\smTEXT}% 
                           \loadviiptfonts\viipttrue\fi 
                           \ifvpt\else\xdef\f@ntsize{\vsTEXT}% 
                           \def\s@ze{5}% 
                           \loadvptfonts\vpttrue\fi 
                      \else\def\TEXT{ten}% 
                           \def\smTEXT{seven}% 
                           \def\vsTEXT{five}% 
                           \ifxpt\else\xdef\f@ntsize{\TEXT}% 
                           \loadxptfonts\xpttrue\fi 
                           \ifviipt\else\xdef\f@ntsize{\smTEXT}% 
                           \def\s@ze{7}% 
                           \loadviiptfonts\viipttrue\fi 
                           \ifvpt\else\xdef\f@ntsize{\vsTEXT}% 
                           \def\s@ze{5}% 
                           \loadvptfonts\vpttrue\fi 
                \ifnum#1=10\else 
                \message{Text size should be 12, 11 or 10 point 
                              will now default to 10pt}\fi 
                \fi\fi\textfonts} 
% 
% Small sized material in 10pt, 9pt or 8pt 
% 
\def\smallsize#1#2{\def\smallb@seline{#2}% 
                 \ifnum#1=10\def\SMALL{ten}% 
                           \def\smSMALL{seven}% 
                           \def\vsSMALL{five}% 
                           \ifxpt\else\xdef\f@ntsize{\SMALL}% 
                           \loadxptfonts\xpttrue\fi 
                           \ifviipt\else\xdef\f@ntsize{\smSMALL}% 
                           \def\s@ze{7}% 
                           \loadviiptfonts\viipttrue\fi 
                           \ifvpt\else\xdef\f@ntsize{\vsSMALL}% 
                           \def\s@ze{5}% 
                           \loadvptfonts\vpttrue\fi 
                       \else 
                 \ifnum#1=9\def\SMALL{nine}% 
                           \def\smSMALL{six}% 
                           \def\vsSMALL{five}% 
                           \ifixpt\else\xdef\f@ntsize{\SMALL}% 
                           \def\s@ze{9}% 
                           \loadsmallfonts\ixpttrue\fi 
                           \ifvipt\else\xdef\f@ntsize{\smSMALL}% 
                           \def\s@ze{6}% 
                           \loadviptfonts\vipttrue\fi 
                           \ifvpt\else\xdef\f@ntsize{\vsSMALL}% 
                           \def\s@ze{5}% 
                           \loadvptfonts\vpttrue\fi 
                       \else 
                           \def\SMALL{eight}% 
                           \def\smSMALL{six}% 
                           \def\vsSMALL{five}% 
                           \ifviiipt\else\xdef\f@ntsize{\SMALL}% 
                           \def\s@ze{8}% 
                           \loadsmallfonts\viiipttrue\fi 
                           \ifvipt\else\xdef\f@ntsize{\smSMALL}% 
                           \def\s@ze{6}% 
                           \loadviptfonts\vipttrue\fi 
                           \ifvpt\else\xdef\f@ntsize{\vsSMALL}% 
                           \def\s@ze{5}% 
                           \loadvptfonts\vpttrue\fi 
                 \ifnum#1=8\else\message{Small size should be 10, 9 or  
                            8 point will now default to 8pt}\fi 
                \fi\fi\smallfonts} 
\def\F@nt{\expandafter\font\csname} 
\def\Sk@w{\expandafter\skewchar\csname} 
\def\@nd{\endcsname} 
\def\@step#1{ scaled \magstep#1} 
\def\@half{ scaled \magstephalf} 
\def\@t#1{ at #1pt} 
% 
% For 14, 17 and 20 point fonts use \loadheadfonts 
% 
\def\loadheadfonts{\bigf@nts 
\F@nt \f@ntsize bdi\@nd=cmmib10 \@t{\s@ze}% 
\Sk@w \f@ntsize bdi\@nd='177 
\F@nt \f@ntsize bsy\@nd=cmbsy10 \@t{\s@ze}% 
\Sk@w \f@ntsize bsy\@nd='60 
\F@nt \f@ntsize bss\@nd=cmssbx10 \@t{\s@ze}} 
% 
% For 12 point fonts use \loadxiiptfonts 
% 
\def\loadxiiptfonts{\bigf@nts 
\F@nt \f@ntsize bdi\@nd=cmmib10 \@step{\m@g}% 
\Sk@w \f@ntsize bdi\@nd='177 
\F@nt \f@ntsize bsy\@nd=cmbsy10 \@step{\m@g}% 
\Sk@w \f@ntsize bsy\@nd='60 
\F@nt \f@ntsize bss\@nd=cmssbx10 \@step{\m@g}} 
% 
% For 11 point fonts use \loadxiptfonts 
% 
\def\loadxiptfonts{% 
\font\elevenrm=cmr10 \@half 
\font\eleveni=cmmi10 \@half 
\skewchar\eleveni='177 
\font\elevensy=cmsy10 \@half 
\skewchar\elevensy='60 
\font\elevenex=cmex10 \@half 
\font\elevenit=cmti10 \@half 
\font\elevensl=cmsl10 \@half 
\font\elevenbf=cmbx10 \@half 
\font\eleventt=cmtt10 \@half 
\ifams\font\elevenmsa=msam10 \@half 
\font\elevenmsb=msbm10 \@half\else\fi 
\font\elevenbdi=cmmib10 \@half 
\skewchar\elevenbdi='177 
\font\elevenbsy=cmbsy10 \@half 
\skewchar\elevenbsy='60 
\font\elevenbss=cmssbx10 \@half} 
% 
% For 10 point fonts use \loadxptfonts 
% 
\def\loadxptfonts{% 
\font\tenbdi=cmmib10 
\skewchar\tenbdi='177 
\font\tenbsy=cmbsy10  
\skewchar\tenbsy='60 
\ifams\font\tenmsa=msam10  
\font\tenmsb=msbm10\else\fi 
\font\tenbss=cmssbx10}%  
% 
% For 8 and 9 point fonts use \loadsmallfonts 
% 
\def\loadsmallfonts{\smallf@nts 
\ifams 
\F@nt \f@ntsize ex\@nd=cmex\s@ze 
\else 
\F@nt \f@ntsize ex\@nd=cmex10\fi 
\F@nt \f@ntsize it\@nd=cmti\s@ze 
\F@nt \f@ntsize sl\@nd=cmsl\s@ze 
\F@nt \f@ntsize tt\@nd=cmtt\s@ze} 
% 
% For 7 point fonts use \loadviiptfonts 
% 
\def\loadviiptfonts{% 
\font\sevenit=cmti7 
\font\sevensl=cmsl8 at 7pt 
\ifams\font\sevenmsa=msam7  
\font\sevenmsb=msbm7 
\font\sevenex=cmex7 
\font\sevenbsy=cmbsy7 
\font\sevenbdi=cmmib7\else 
\font\sevenex=cmex10 
\font\sevenbsy=cmbsy10 at 7pt 
\font\sevenbdi=cmmib10 at 7pt\fi 
\skewchar\sevenbsy='60 
\skewchar\sevenbdi='177 
\font\sevenbss=cmssbx10 at 7pt}%  
% 
%  For 6 point fonts use \loadviptfonts 
% 
\def\loadviptfonts{\smallf@nts 
\ifams\font\sixex=cmex7 at 6pt\else 
\font\sixex=cmex10\fi 
\font\sixit=cmti7 at 6pt} 
% 
% For 5 point fonts use \loadvptfonts 
% 
\def\loadvptfonts{% 
\font\fiveit=cmti7 at 5pt 
\ifams\font\fiveex=cmex7 at 5pt 
\font\fivebdi=cmmib5 
\font\fivebsy=cmbsy5 
\font\fivemsa=msam5  
\font\fivemsb=msbm5\else 
\font\fiveex=cmex10 
\font\fivebdi=cmmib10 at 5pt 
\font\fivebsy=cmbsy10 at 5pt\fi 
\skewchar\fivebdi='177 
\skewchar\fivebsy='60 
\font\fivebss=cmssbx10 at 5pt} 
\def\bigf@nts{% 
\F@nt \f@ntsize rm\@nd=cmr10 \@step{\m@g}% 
\F@nt \f@ntsize i\@nd=cmmi10 \@step{\m@g}% 
\Sk@w \f@ntsize i\@nd='177 
\F@nt \f@ntsize sy\@nd=cmsy10 \@step{\m@g}% 
\Sk@w \f@ntsize sy\@nd='60 
\F@nt \f@ntsize ex\@nd=cmex10 \@step{\m@g}% 
\F@nt \f@ntsize it\@nd=cmti10 \@step{\m@g}% 
\F@nt \f@ntsize sl\@nd=cmsl10 \@step{\m@g}% 
\F@nt \f@ntsize bf\@nd=cmbx10 \@step{\m@g}% 
\F@nt \f@ntsize tt\@nd=cmtt10 \@step{\m@g}% 
\ifams 
\F@nt \f@ntsize msa\@nd=msam10 \@step{\m@g}% 
\F@nt \f@ntsize msb\@nd=msbm10 \@step{\m@g}\else\fi} 
\def\smallf@nts{% 
\F@nt \f@ntsize rm\@nd=cmr\s@ze 
\F@nt \f@ntsize i\@nd=cmmi\s@ze  
\Sk@w \f@ntsize i\@nd='177 
\F@nt \f@ntsize sy\@nd=cmsy\s@ze 
\Sk@w \f@ntsize sy\@nd='60 
\F@nt \f@ntsize bf\@nd=cmbx\s@ze  
\ifams 
\F@nt \f@ntsize bdi\@nd=cmmib\s@ze  
\F@nt \f@ntsize bsy\@nd=cmbsy\s@ze  
\F@nt \f@ntsize msa\@nd=msam\s@ze  
\F@nt \f@ntsize msb\@nd=msbm\s@ze 
\else 
\F@nt \f@ntsize bdi\@nd=cmmib10 \@t{\s@ze}%  
\F@nt \f@ntsize bsy\@nd=cmbsy10 \@t{\s@ze}\fi  
\Sk@w \f@ntsize bdi\@nd='177 
\Sk@w \f@ntsize bsy\@nd='60 
\F@nt \f@ntsize bss\@nd=cmssbx10 \@t{\s@ze}}%  
% 
% Fonts for headings  
% 
\def\headfonts{% 
\textfont0=\csname\HEAD rm\@nd         
\scriptfont0=\csname\smHEAD rm\@nd 
\scriptscriptfont0=\csname\vsHEAD rm\@nd 
\def\rm{\fam0\csname\HEAD rm\@nd 
\def\sc{\csname\smHEAD rm\@nd}}% 
\textfont1=\csname\HEAD i\@nd          
\scriptfont1=\csname\smHEAD i\@nd 
\scriptscriptfont1=\csname\vsHEAD i\@nd 
\textfont2=\csname\HEAD sy\@nd         
\scriptfont2=\csname\smHEAD sy\@nd 
\scriptscriptfont2=\csname\vsHEAD sy\@nd 
\textfont3=\csname\HEAD ex\@nd         
\scriptfont3=\csname\smHEAD ex\@nd 
\scriptscriptfont3=\csname\smHEAD ex\@nd 
\textfont\itfam=\csname\HEAD it\@nd    
\scriptfont\itfam=\csname\smHEAD it\@nd 
\scriptscriptfont\itfam=\csname\vsHEAD it\@nd 
\def\it{\fam\itfam\csname\HEAD it\@nd 
\def\sc{\csname\smHEAD it\@nd}}% 
\textfont\slfam=\csname\HEAD sl\@nd    
\def\sl{\fam\slfam\csname\HEAD sl\@nd 
\def\sc{\csname\smHEAD sl\@nd}}% 
\textfont\bffam=\csname\HEAD bf\@nd    
\scriptfont\bffam=\csname\smHEAD bf\@nd 
\scriptscriptfont\bffam=\csname\vsHEAD bf\@nd 
\def\bf{\fam\bffam\csname\HEAD bf\@nd 
\def\sc{\csname\smHEAD bf\@nd}}% 
\textfont\ttfam=\csname\HEAD tt\@nd    
\def\tt{\fam\ttfam\csname\HEAD tt\@nd}% 
\textfont\bdifam=\csname\HEAD bdi\@nd  
\scriptfont\bdifam=\csname\smHEAD bdi\@nd 
\scriptscriptfont\bdifam=\csname\vsHEAD bdi\@nd 
\def\bdi{\fam\bdifam\csname\HEAD bdi\@nd}% 
\textfont\bsyfam=\csname\HEAD bsy\@nd  
\scriptfont\bsyfam=\csname\smHEAD bsy\@nd 
\def\bsy{\fam\bsyfam\csname\HEAD bsy\@nd}% 
\textfont\bssfam=\csname\HEAD bss\@nd  
\scriptfont\bssfam=\csname\smHEAD bss\@nd 
\scriptscriptfont\bssfam=\csname\vsHEAD bss\@nd 
\def\bss{\fam\bssfam\csname\HEAD bss\@nd}% 
\ifams 
\textfont\msafam=\csname\HEAD msa\@nd  
\scriptfont\msafam=\csname\smHEAD msa\@nd 
\scriptscriptfont\msafam=\csname\vsHEAD msa\@nd 
\textfont\msbfam=\csname\HEAD msb\@nd  
\scriptfont\msbfam=\csname\smHEAD msb\@nd 
\scriptscriptfont\msbfam=\csname\vsHEAD msb\@nd 
\else\fi 
\normalbaselineskip=\headb@seline pt% 
\setbox\strutbox=\hbox{\vrule height.7\normalbaselineskip  
depth.3\baselineskip width0pt}% 
\def\sc{\csname\smHEAD rm\@nd}\normalbaselines\bf} 
% 
% Fonts for text 
% 
\def\textfonts{% 
\textfont0=\csname\TEXT rm\@nd         
\scriptfont0=\csname\smTEXT rm\@nd 
\scriptscriptfont0=\csname\vsTEXT rm\@nd 
\def\rm{\fam0\csname\TEXT rm\@nd 
\def\sc{\csname\smTEXT rm\@nd}}% 
\textfont1=\csname\TEXT i\@nd          
\scriptfont1=\csname\smTEXT i\@nd 
\scriptscriptfont1=\csname\vsTEXT i\@nd 
\textfont2=\csname\TEXT sy\@nd         
\scriptfont2=\csname\smTEXT sy\@nd 
\scriptscriptfont2=\csname\vsTEXT sy\@nd 
\textfont3=\csname\TEXT ex\@nd         
\scriptfont3=\csname\smTEXT ex\@nd 
\scriptscriptfont3=\csname\smTEXT ex\@nd 
\textfont\itfam=\csname\TEXT it\@nd    
\scriptfont\itfam=\csname\smTEXT it\@nd 
\scriptscriptfont\itfam=\csname\vsTEXT it\@nd 
\def\it{\fam\itfam\csname\TEXT it\@nd 
\def\sc{\csname\smTEXT it\@nd}}% 
\textfont\slfam=\csname\TEXT sl\@nd    
\def\sl{\fam\slfam\csname\TEXT sl\@nd 
\def\sc{\csname\smTEXT sl\@nd}}% 
\textfont\bffam=\csname\TEXT bf\@nd    
\scriptfont\bffam=\csname\smTEXT bf\@nd 
\scriptscriptfont\bffam=\csname\vsTEXT bf\@nd 
\def\bf{\fam\bffam\csname\TEXT bf\@nd 
\def\sc{\csname\smTEXT bf\@nd}}% 
\textfont\ttfam=\csname\TEXT tt\@nd    
\def\tt{\fam\ttfam\csname\TEXT tt\@nd}% 
\textfont\bdifam=\csname\TEXT bdi\@nd  
\scriptfont\bdifam=\csname\smTEXT bdi\@nd 
\scriptscriptfont\bdifam=\csname\vsTEXT bdi\@nd 
\def\bdi{\fam\bdifam\csname\TEXT bdi\@nd}% 
\textfont\bsyfam=\csname\TEXT bsy\@nd  
\scriptfont\bsyfam=\csname\smTEXT bsy\@nd 
\def\bsy{\fam\bsyfam\csname\TEXT bsy\@nd}% 
\textfont\bssfam=\csname\TEXT bss\@nd  
\scriptfont\bssfam=\csname\smTEXT bss\@nd 
\scriptscriptfont\bssfam=\csname\vsTEXT bss\@nd 
\def\bss{\fam\bssfam\csname\TEXT bss\@nd}% 
\ifams 
\textfont\msafam=\csname\TEXT msa\@nd  
\scriptfont\msafam=\csname\smTEXT msa\@nd 
\scriptscriptfont\msafam=\csname\vsTEXT msa\@nd 
\textfont\msbfam=\csname\TEXT msb\@nd  
\scriptfont\msbfam=\csname\smTEXT msb\@nd 
\scriptscriptfont\msbfam=\csname\vsTEXT msb\@nd 
\else\fi 
\normalbaselineskip=\textb@seline pt 
\setbox\strutbox=\hbox{\vrule height.7\normalbaselineskip  
depth.3\baselineskip width0pt}% 
\everymath{}% 
\def\sc{\csname\smTEXT rm\@nd}\normalbaselines\rm} 
% 
% Fonts for small material (captions, footnotes etc) 
% 
\def\smallfonts{% 
\textfont0=\csname\SMALL rm\@nd         
\scriptfont0=\csname\smSMALL rm\@nd 
\scriptscriptfont0=\csname\vsSMALL rm\@nd 
\def\rm{\fam0\csname\SMALL rm\@nd 
\def\sc{\csname\smSMALL rm\@nd}}% 
\textfont1=\csname\SMALL i\@nd          
\scriptfont1=\csname\smSMALL i\@nd 
\scriptscriptfont1=\csname\vsSMALL i\@nd 
\textfont2=\csname\SMALL sy\@nd         
\scriptfont2=\csname\smSMALL sy\@nd 
\scriptscriptfont2=\csname\vsSMALL sy\@nd 
\textfont3=\csname\SMALL ex\@nd         
\scriptfont3=\csname\smSMALL ex\@nd 
\scriptscriptfont3=\csname\smSMALL ex\@nd 
\textfont\itfam=\csname\SMALL it\@nd    
\scriptfont\itfam=\csname\smSMALL it\@nd 
\scriptscriptfont\itfam=\csname\vsSMALL it\@nd 
\def\it{\fam\itfam\csname\SMALL it\@nd 
\def\sc{\csname\smSMALL it\@nd}}% 
\textfont\slfam=\csname\SMALL sl\@nd    
\def\sl{\fam\slfam\csname\SMALL sl\@nd 
\def\sc{\csname\smSMALL sl\@nd}}% 
\textfont\bffam=\csname\SMALL bf\@nd    
\scriptfont\bffam=\csname\smSMALL bf\@nd 
\scriptscriptfont\bffam=\csname\vsSMALL bf\@nd 
\def\bf{\fam\bffam\csname\SMALL bf\@nd 
\def\sc{\csname\smSMALL bf\@nd}}% 
\textfont\ttfam=\csname\SMALL tt\@nd    
\def\tt{\fam\ttfam\csname\SMALL tt\@nd}% 
\textfont\bdifam=\csname\SMALL bdi\@nd  
\scriptfont\bdifam=\csname\smSMALL bdi\@nd 
\scriptscriptfont\bdifam=\csname\vsSMALL bdi\@nd 
\def\bdi{\fam\bdifam\csname\SMALL bdi\@nd}% 
\textfont\bsyfam=\csname\SMALL bsy\@nd  
\scriptfont\bsyfam=\csname\smSMALL bsy\@nd 
\def\bsy{\fam\bsyfam\csname\SMALL bsy\@nd}% 
\textfont\bssfam=\csname\SMALL bss\@nd  
\scriptfont\bssfam=\csname\smSMALL bss\@nd 
\scriptscriptfont\bssfam=\csname\vsSMALL bss\@nd 
\def\bss{\fam\bssfam\csname\SMALL bss\@nd}% 
\ifams 
\textfont\msafam=\csname\SMALL msa\@nd  
\scriptfont\msafam=\csname\smSMALL msa\@nd 
\scriptscriptfont\msafam=\csname\vsSMALL msa\@nd 
\textfont\msbfam=\csname\SMALL msb\@nd  
\scriptfont\msbfam=\csname\smSMALL msb\@nd 
\scriptscriptfont\msbfam=\csname\vsSMALL msb\@nd 
\else\fi 
\normalbaselineskip=\smallb@seline pt% 
\setbox\strutbox=\hbox{\vrule height.7\normalbaselineskip  
depth.3\baselineskip width0pt}% 
\everymath{}% 
\def\sc{\csname\smSMALL rm\@nd}\normalbaselines\rm}% 
\everydisplay{\indenteddisplay 
   \gdef\labeltype{\eqlabel}}% 
% 
%%%%%%%%%%%%%%%%%%%%%%%%%%%%%%%%%%%%%%%%%%%%%%%%%%%%%%%%%%% 
%                                                         % 
%  Macros to define extra maths symbols                   % 
%                                                         % 
%%%%%%%%%%%%%%%%%%%%%%%%%%%%%%%%%%%%%%%%%%%%%%%%%%%%%%%%%%% 
% 
\def\hexnumber@#1{\ifcase#1 0\or 1\or 2\or 3\or 4\or 5\or 6\or 7\or 8\or 
 9\or A\or B\or C\or D\or E\or F\fi} 
\edef\bffam@{\hexnumber@\bffam} 
\edef\bdifam@{\hexnumber@\bdifam} 
\edef\bsyfam@{\hexnumber@\bsyfam} 
\def\undefine#1{\let#1\undefined} 
\def\newsymbol#1#2#3#4#5{\let\next@\relax 
 \ifnum#2=\thr@@\let\next@\bdifam@\else 
 \ifams 
 \ifnum#2=\@ne\let\next@\msafam@\else 
 \ifnum#2=\tw@\let\next@\msbfam@\fi\fi 
 \fi\fi 
 \mathchardef#1="#3\next@#4#5} 
\def\mathhexbox@#1#2#3{\relax 
 \ifmmode\mathpalette{}{\m@th\mathchar"#1#2#3}% 
 \else\leavevmode\hbox{$\m@th\mathchar"#1#2#3$}\fi} 

\def\bi#1{{\fam\bdifam\relax#1}} 
% 
% If file amsmacro is not in current directory 
% or somewhere with set path add path before 
% file name in following line 
% 
\ifams\input amsmacro\fi 
% 
% Bold italic Greek characters 
% 
\newsymbol\bitGamma 3000 
\newsymbol\bitDelta 3001 
\newsymbol\bitTheta 3002 
\newsymbol\bitLambda 3003 
\newsymbol\bitXi 3004 
\newsymbol\bitPi 3005 
\newsymbol\bitSigma 3006 
\newsymbol\bitUpsilon 3007 
\newsymbol\bitPhi 3008 
\newsymbol\bitPsi 3009 
\newsymbol\bitOmega 300A 
\newsymbol\balpha 300B 
\newsymbol\bbeta 300C 
\newsymbol\bgamma 300D 
\newsymbol\bdelta 300E 
\newsymbol\bepsilon 300F 
\newsymbol\bzeta 3010 
\newsymbol\bfeta 3011 
\newsymbol\btheta 3012 
\newsymbol\biota 3013 
\newsymbol\bkappa 3014 
\newsymbol\blambda 3015 
\newsymbol\bmu 3016 
\newsymbol\bnu 3017 
\newsymbol\bxi 3018 
\newsymbol\bpi 3019 
\newsymbol\brho 301A 
\newsymbol\bsigma 301B 
\newsymbol\btau 301C 
\newsymbol\bupsilon 301D 
\newsymbol\bphi 301E 
\newsymbol\bchi 301F 
\newsymbol\bpsi 3020 
\newsymbol\bomega 3021 
\newsymbol\bvarepsilon 3022 
\newsymbol\bvartheta 3023 
\newsymbol\bvaromega 3024 
\newsymbol\bvarrho 3025 
\newsymbol\bvarzeta 3026 
\newsymbol\bvarphi 3027 
\newsymbol\bpartial 3040 
\newsymbol\bell 3060 
\newsymbol\bimath 307B 
\newsymbol\bjmath 307C 
\mathchardef\binfty "0\bsyfam@31 
\mathchardef\bnabla "0\bsyfam@72 
\mathchardef\bdot "2\bsyfam@01 
\mathchardef\bGamma "0\bffam@00 
\mathchardef\bDelta "0\bffam@01 
\mathchardef\bTheta "0\bffam@02 
\mathchardef\bLambda "0\bffam@03 
\mathchardef\bXi "0\bffam@04 
\mathchardef\bPi "0\bffam@05 
\mathchardef\bSigma "0\bffam@06 
\mathchardef\bUpsilon "0\bffam@07 
\mathchardef\bPhi "0\bffam@08 
\mathchardef\bPsi "0\bffam@09 
\mathchardef\bOmega "0\bffam@0A 
\mathchardef\itGamma "0100 
\mathchardef\itDelta "0101 
\mathchardef\itTheta "0102 
\mathchardef\itLambda "0103 
\mathchardef\itXi "0104 
\mathchardef\itPi "0105 
\mathchardef\itSigma "0106 
\mathchardef\itUpsilon "0107 
\mathchardef\itPhi "0108 
\mathchardef\itPsi "0109 
\mathchardef\itOmega "010A 
\mathchardef\Gamma "0000 
\mathchardef\Delta "0001 
\mathchardef\Theta "0002 
\mathchardef\Lambda "0003 
\mathchardef\Xi "0004 
\mathchardef\Pi "0005 
\mathchardef\Sigma "0006 
\mathchardef\Upsilon "0007 
\mathchardef\Phi "0008 
\mathchardef\Psi "0009 
\mathchardef\Omega "000A 
% 
% Counter definitions 
% 
\newcount\firstpage  \firstpage=1  % start page no 
\newcount\jnl                      % journal no 
\newcount\secno                    % section number 
\newcount\subno                    % number of subsection 
\newcount\subsubno                 % number of subsubsection 
\newcount\appno                    % appendix number 
\newcount\tabno                    % table number 
\newcount\figno                    % figure number 
\newcount\countno                  % equation numbers 
\newcount\refno                    % reference number 
\newcount\eqlett     \eqlett=97    % equation letter 
\newif\ifletter 
\newif\ifwide 
\newif\ifnotfull 
\newif\ifaligned 
\newif\ifnumbysec   
\newif\ifappendix 
\newif\ifnumapp 
\newif\ifssf 
\newif\ifppt 
\newdimen\t@bwidth 
\newdimen\c@pwidth 
\newdimen\digitwidth                    %character width 
\newdimen\argwidth                      %argument width 
\newdimen\secindent    \secindent=5pc   %indentation of maths  
\newdimen\textind    \textind=16pt      %indentation of text 
\newdimen\tempval                       %temporary value 
\newskip\beforesecskip 
\def\beforesecspace{\vskip\beforesecskip\relax} 
\newskip\beforesubskip 
\def\beforesubspace{\vskip\beforesubskip\relax} 
\newskip\beforesubsubskip 
\def\beforesubsubspace{\vskip\beforesubsubskip\relax} 
\newskip\secskip 
\def\secspace{\vskip\secskip\relax} 
\newskip\subskip 
\def\subspace{\vskip\subskip\relax} 
\newskip\insertskip 
\def\insertspace{\vskip\insertskip\relax} 
\def\sp@ce{\ifx\next*\let\next=\@ssf 
               \else\let\next=\@nossf\fi\next} 
\def\@ssf#1{\nobreak\secspace\global\ssftrue\nobreak} 
\def\@nossf{\nobreak\secspace\nobreak\noindent\ignorespaces} 
\def\subsp@ce{\ifx\next*\let\next=\@sssf 
               \else\let\next=\@nosssf\fi\next} 
\def\@sssf#1{\nobreak\subspace\global\ssftrue\nobreak} 
\def\@nosssf{\nobreak\subspace\nobreak\noindent\ignorespaces} 
\beforesecskip=24pt plus12pt minus8pt 
\beforesubskip=12pt plus6pt minus4pt 
\beforesubsubskip=12pt plus6pt minus4pt 
\secskip=12pt plus 2pt minus 2pt 
\subskip=6pt plus3pt minus2pt 
\insertskip=18pt plus6pt minus6pt% 
\fontdimen16\tensy=2.7pt 
\fontdimen17\tensy=2.7pt 
% 
% Labels etc for cross referencing macros 
% 
\def\eqlabel{(\ifappendix\applett 
               \ifnumbysec\ifnum\secno>0 \the\secno\fi.\fi 
               \else\ifnumbysec\the\secno.\fi\fi\the\countno)} 
\def\seclabel{\ifappendix\ifnumapp\else\applett\fi 
    \ifnum\secno>0 \the\secno 
    \ifnumbysec\ifnum\subno>0.\the\subno\fi\fi\fi 
    \else\the\secno\fi\ifnum\subno>0.\the\subno 
         \ifnum\subsubno>0.\the\subsubno\fi\fi} 
\def\tablabel{\ifappendix\applett\fi\the\tabno} 
\def\figlabel{\ifappendix\applett\fi\the\figno} 
\def\gac{\global\advance\countno by 1} 
% 
% Redefinition of footnote macros to lose rule and remove indentation 
% 
 
\def\vfootnote#1{\insert\footins\bgroup 
\interlinepenalty=\interfootnotelinepenalty 
\splittopskip=\ht\strutbox % top baseline for broken footnotes 
\splitmaxdepth=\dp\strutbox \floatingpenalty=20000 
\leftskip=0pt \rightskip=0pt \spaceskip=0pt \xspaceskip=0pt% 
\noindent\smallfonts\rm #1\ \ignorespaces\footstrut\futurelet\next\fo@t} 
% 
% Redefinition of endinsert to give more controllable 
% space around  tables and figures 
% 
\def\endinsert{\egroup 
    \if@mid \dimen@=\ht0 \advance\dimen@ by\dp0 
       \advance\dimen@ by12\p@ \advance\dimen@ by\pagetotal 
       \ifdim\dimen@>\pagegoal \@midfalse\p@gefalse\fi\fi 
    \if@mid \insertspace \box0 \par \ifdim\lastskip<\insertskip 
    \removelastskip \penalty-200 \insertspace \fi 
    \else\insert\topins{\penalty100 
       \splittopskip=0pt \splitmaxdepth=\maxdimen  
       \floatingpenalty=0 
       \ifp@ge \dimen@=\dp0 
       \vbox to\vsize{\unvbox0 \kern-\dimen@}% 
       \else\box0\nobreak\insertspace\fi}\fi\endgroup}    
% 
% special macros for display equations 
% 
% for indentation of turned over lines in mathematics 
% 
\def\ind{\hbox to \secindent{\hfill}} 
% 
% for turned over equals sign to left of maths indent 
% 

% 
% for other signs to left of maths indent 
% 
 
% 
% displayed equation indented  
% 
\def\indeqn#1{\alignedfalse\displ@y\halign{\hbox to \displaywidth 
    {$\ind\@lign\displaystyle##\hfil$}\crcr #1\crcr}} 
% 
% displayed equation indented with alignments 
% 
\def\indalign#1{\alignedtrue\displ@y \tabskip=0pt  
  \halign to\displaywidth{\ind$\@lign\displaystyle{##}$\tabskip=0pt 
    &$\@lign\displaystyle{{}##}$\hfill\tabskip=\centering 
    &\llap{$\@lign\hbox{\rm##}$}\tabskip=0pt\crcr 
    #1\crcr}} 
\def\fl{{\hskip-\secindent}} 
\def\indenteddisplay#1$${\indispl@y{#1 }} 
\def\indispl@y#1{\disptest#1\eqalignno\eqalignno\disptest} 
\def\disptest#1\eqalignno#2\eqalignno#3\disptest{% 
    \ifx#3\eqalignno 
    \indalign#2% 
    \else\indeqn{#1}\fi$$} 
% 
% Roman small caps (if in Roman \sc gives small caps) 
% 
 
% 
% Italic small caps (if in italic \sc gives italic small caps) 
% 
 
% 
% Bold small caps (if in bold \sc gives bold small caps) 
% 
 
% 
% Small caps in maths 
% 
 
% 
% Miscellaneous definitions 
% 

\def\ns{\noalign{\vskip-3pt}}

% 
 
% 
% Bold h bar 
% 
\def\bhbar{\rlap{\kern1pt\raise.4ex\hbox{\bf\char'40}}\bi{h}} 
\def\case#1#2{{\textstyle{#1\over#2}}}

\def\frac#1#2{{#1\over#2}} 
\ifams 
\def\lap{\lesssim} 
\def\gap{\gtrsim}

\else

\def\gap{\;\lower3pt\hbox{$\buildrel > \over \sim$}\;}% 
\def\lap{\;\lower3pt\hbox{$\buildrel < \over \sim$}\;}\fi 
 
\chardef\ii="10 
\def\tqs{\hbox to 25pt{\hfil}}

\def\Bbbone{1\kern-.22em {\rm l}} 
% 
% Primes to display summations and products  
% which also have sub or superscripts 
% 
\def\rp{\raise8pt\hbox{$\scriptstyle\prime$}} 
% 
% then use \sum^{...}_{...}\rp or \prod^{...}_{...}\rp. 
% 
% Shadow brackets 
% 
% Single brackets for normal size only 
% 

% 
% Variable size for display style 
% 
\def\[#1\]{\setbox0=\hbox{$\dsty#1$}\argwidth=\wd0 
    \setbox0=\hbox{$\left[\box0\right]$}\advance\argwidth by -\wd0 
    \left[\kern.3\argwidth\box0\kern.3\argwidth\right]} 
% 
% Variable size for text style 
% 
\def\lsb#1\rsb{\setbox0=\hbox{$#1$}\argwidth=\wd0 
    \setbox0=\hbox{$\left[\box0\right]$}\advance\argwidth by -\wd0 
    \left[\kern.3\argwidth\box0\kern.3\argwidth\right]} 
% 
 
% 
% Square for end of theorems 
% 
 
% 
\def\pt(#1){({\it #1\/})} 
\let\dsty=\displaystyle

% 
% Definition for Nuclear Physics Keyword abstract 
% 
\def\reactions#1{\vskip 12pt plus2pt minus2pt%              
\vbox{\hbox{\kern\secindent\vrule\kern12pt% 
\vbox{\kern0.5pt\vbox{\hsize=24pc\parindent=0pt\smallfonts\rm NUCLEAR  
REACTIONS\strut\quad #1\strut}\kern0.5pt}\kern12pt\vrule}}} 
% 
% Definition for slashed characters 
% 
\def\slashchar#1{\setbox0=\hbox{$#1$}\dimen0=\wd0% 
\setbox1=\hbox{/}\dimen1=\wd1% 
\ifdim\dimen0>\dimen1%                         
\rlap{\hbox to \dimen0{\hfil/\hfil}}#1\else                                         
\rlap{\hbox to \dimen1{\hfil$#1$\hfil}}/\fi} 
% 
% Redefine \textindent for use in \item 
% 
\def\textindent#1{\noindent\hbox to \parindent{#1\hss}\ignorespaces} 
% 
% Symbols and curves for use in figure captions 
% 
\def\opencirc{\raise1pt\hbox{$\scriptstyle{\bigcirc}$}} 
\def\fullcirc{\hbox{\headfonts\rm$\scriptstyle\bullet$}} 
\ifams 
\def\opensqr{\hbox{$\square$}} 
\def\opentri{\hbox{$\vartriangle$}} 
\def\opentridown{\hbox{$\triangledown$}} 
\def\opendiamond{\hbox{$\lozenge$}}

\else 
\def\opensqr{\vbox{\hrule height.4pt\hbox{\vrule width.4pt height3.5pt 
    \kern3.5pt\vrule width.4pt}\hrule height.4pt}} 
\def\opentri{\hbox{$\scriptstyle\bigtriangleup$}} 
\def\opentridown{\raise1pt\hbox{$\scriptstyle\bigtriangledown$}} 
\def\opendiamond{\hbox{$\diamond$}} 
 
           %  These produce the 
                   %  equivalent open character 
           %  to be filled in. 
\fi

% 
% Redefinition of \cases 
% 
\def\m@th{\mathsurround=0pt} 
% 
% Displaystyle now used for first term 
% 
\def\cases#1{% 
\left\{\,\vcenter{\normalbaselines\openup1\jot\m@th% 
     \ialign{$\displaystyle##\hfil$&\rm\tqs##\hfil\crcr#1\crcr}}\right.}% 
% 
% Original version of cases now called \oldcases 
% 
\def\oldcases#1{\left\{\,\vcenter{\normalbaselines\m@th 
    \ialign{$##\hfil$&\rm\quad##\hfil\crcr#1\crcr}}\right.} 
% 
% Cases with number at end each line (using automatic numbering) 
% 
\def\numcases#1{\left\{\,\vcenter{\baselineskip=15pt\m@th% 
     \ialign{$\displaystyle##\hfil$&\rm\tqs##\hfil 
     \crcr#1\crcr}}\right.\hfill 
     \vcenter{\baselineskip=15pt\m@th% 
     \ialign{\rlap{$\phantom{\displaystyle##\hfil}$}\tabskip=0pt&\en 
     \rlap{\phantom{##\hfil}}\crcr#1\crcr}}} 
\def\ptnumcases#1{\left\{\,\vcenter{\baselineskip=15pt\m@th% 
     \ialign{$\displaystyle##\hfil$&\rm\tqs##\hfil 
     \crcr#1\crcr}}\right.\hfill 
     \vcenter{\baselineskip=15pt\m@th% 
     \ialign{\rlap{$\phantom{\displaystyle##\hfil}$}\tabskip=0pt&\enpt 
     \rlap{\phantom{##\hfil}}\crcr#1\crcr}}\global\eqlett=97 
     \global\advance\countno by 1} 
% 
% for equation numbers instead of \eqno 
% 
\def\eq(#1){\ifaligned\@mp(#1)\else\hfill\llap{{\rm (#1)}}\fi} 
\def\ceq(#1){\ns\ns\ifaligned\@mp\fi\eq(#1)\cr\ns\ns} 
\def\eqpt(#1#2){\ifaligned\@mp(#1{\it #2\/}) 
                    \else\hfill\llap{{\rm (#1{\it #2\/})}}\fi} 
 
% 
% Automatic numbering of equations 
% 
\countno=1 
\def\eqnobysec{\numbysectrue} 
\def\aleq{&\rm(\ifappendix\applett 
               \ifnumbysec\ifnum\secno>0 \the\secno\fi.\fi 
               \else\ifnumbysec\the\secno.\fi\fi\the\countno} 
\def\noaleq{\hfill\llap\bgroup\rm(\ifappendix\applett 
               \ifnumbysec\ifnum\secno>0 \the\secno\fi.\fi 
               \else\ifnumbysec\the\secno.\fi\fi\the\countno} 
\def\@mp{&} 
\def\en{\ifaligned\aleq)\else\noaleq)\egroup\fi\gac} 
\def\cen{\ns\ns\ifaligned\@mp\fi\en\cr\ns\ns} 
\def\enpt{\ifaligned\aleq{\it\char\the\eqlett})\else 
    \noaleq{\it\char\the\eqlett})\egroup\fi 
    \global\advance\eqlett by 1} 
\def\endpt{\ifaligned\aleq{\it\char\the\eqlett})\else 
    \noaleq{\it\char\the\eqlett})\egroup\fi 
    \global\eqlett=97\gac} 
% 
% abbreviations for Institute of Physics Publishing journals 
% 

\def\JPA{{\it J. Phys. A: Math. Gen.}} 
        %1968-87 
   %1988 and onwards 
     %1968--1988 
        %1989 and onwards 

           %1975--1988 
     %1989 and onwards 
 
                 %1990 and onwards 

% 
% Other commonly quoted journals 
% 

\def\AP{{\it Ann. Phys., Lpz.}}

\def\NP{{\it Nucl. Phys.}} 
 
\def\PR{{\it Phys. Rev.}} 
\def\PRL{{\it Phys. Rev. Lett.}}

\def\RMP{{\it Rev. Mod. Phys.}}

\headline={\ifodd\pageno{\ifnum\pageno=\firstpage\hfill 
   \else\rrhead\fi}\else\lrhead\fi} 
\def\rrhead{\textfonts\hskip\secindent\it 
    \shorttitle\hfill\rm\folio} 
\def\lrhead{\textfonts\hbox to\secindent{\rm\folio\hss}% 
    \it\aunames\hss} 
\footline={\ifnum\pageno=\firstpage \hfill\textfonts\rm\folio\fi} 
\def\@rticle#1#2{\vglue.5pc 
    {\parindent=\secindent \bf #1\par} 
     \vskip2.5pc 
    {\exhyphenpenalty=10000\hyphenpenalty=10000 
     \baselineskip=18pt\raggedright\noindent 
     \headfonts\bf#2\par}\futurelet\next\sh@rttitle}% 
\def\title#1{\gdef\shorttitle{#1} 
    \vglue4pc{\exhyphenpenalty=10000\hyphenpenalty=10000  
    \baselineskip=18pt  
    \raggedright\parindent=0pt 
    \headfonts\bf#1\par}\futurelet\next\sh@rttitle}  

\def\article#1#2{\gdef\shorttitle{#2}\@rticle{#1}{#2}}  
\def\review#1{\gdef\shorttitle{#1}% 
    \@rticle{REVIEW \ifpbm\else ARTICLE\fi}{#1}} 
\def\topical#1{\gdef\shorttitle{#1}% 
    \@rticle{TOPICAL REVIEW}{#1}} 
\def\comment#1{\gdef\shorttitle{#1}% 
    \@rticle{COMMENT}{#1}} 
\def\note#1{\gdef\shorttitle{#1}% 
    \@rticle{NOTE}{#1}} 
\def\prelim#1{\gdef\shorttitle{#1}% 
    \@rticle{PRELIMINARY COMMUNICATION}{#1}} 
\def\letter#1{\gdef\shorttitle{Letter to the Editor}% 
     \gdef\aunames{Letter to the Editor} 
     \global\lettertrue\ifnum\jnl=7\global\letterfalse\fi 
     \@rticle{LETTER TO THE EDITOR}{#1}} 
\def\sh@rttitle{\ifx\next[\let\next=\sh@rt 
                \else\let\next=\f@ll\fi\next} 
\def\sh@rt[#1]{\gdef\shorttitle{#1}} 
\def\f@ll{} 
\def\author#1{\ifletter\else\gdef\aunames{#1}\fi\vskip1.5pc 
    {\parindent=\secindent   
     \hang\textfonts   
     \ifppt\bf\else\rm\fi#1\par}   
     \ifppt\bigskip\else\smallskip\fi 
     \futurelet\next\@unames} 
\def\@unames{\ifx\next[\let\next=\short@uthor 
                 \else\let\next=\@uthor\fi\next} 
\def\short@uthor[#1]{\gdef\aunames{#1}} 
\def\@uthor{} 
\def\address#1{{\parindent=\secindent 
    \exhyphenpenalty=10000\hyphenpenalty=10000 
\ifppt\textfonts\else\smallfonts\fi\hang\raggedright\rm#1\par}% 
    \ifppt\bigskip\fi} 
\def\jl#1{\global\jnl=#1} 
\jl{0}% 
\def\journal{\ifnum\jnl=1 J. Phys.\ A: Math.\ Gen.\  
        \else\ifnum\jnl=2 J. Phys.\ B: At.\ Mol.\ Opt.\ Phys.\  
        \else\ifnum\jnl=3 J. Phys.:\ Condens. Matter\  
        \else\ifnum\jnl=4 J. Phys.\ G: Nucl.\ Part.\ Phys.\  
        \else\ifnum\jnl=5 Inverse Problems\  
        \else\ifnum\jnl=6 Class. Quantum Grav.\  
        \else\ifnum\jnl=7 Network\  
        \else\ifnum\jnl=8 Nonlinearity\ 
        \else\ifnum\jnl=9 Quantum Opt.\ 
        \else\ifnum\jnl=10 Waves in Random Media\ 
        \else\ifnum\jnl=11 Pure Appl. Opt.\  
        \else\ifnum\jnl=12 Phys. Med. Biol.\ 
        \else\ifnum\jnl=13 Modelling Simulation Mater.\ Sci.\ Eng.\  
        \else\ifnum\jnl=14 Plasma Phys. Control. Fusion\  
        \else\ifnum\jnl=15 Physiol. Meas.\  
        \else\ifnum\jnl=16 Sov.\ Lightwave Commun.\ 
        \else\ifnum\jnl=17 J. Phys.\ D: Appl.\ Phys.\ 
        \else\ifnum\jnl=18 Supercond.\ Sci.\ Technol.\ 
        \else\ifnum\jnl=19 Semicond.\ Sci.\ Technol.\ 
        \else\ifnum\jnl=20 Nanotechnology\ 
        \else\ifnum\jnl=21 Meas.\ Sci.\ Technol.\  
        \else\ifnum\jnl=22 Plasma Sources Sci.\ Technol.\  
        \else\ifnum\jnl=23 Smart Mater.\ Struct.\  
        \else\ifnum\jnl=24 J.\ Micromech.\ Microeng.\ 
   \else Institute of Physics Publishing\  
   \fi\fi\fi\fi\fi\fi\fi\fi\fi\fi\fi\fi\fi\fi\fi 
   \fi\fi\fi\fi\fi\fi\fi\fi\fi} 
\let\abs=\beginabstract 

\let\endabs=\endabstract 
\def\submitted{\ifppt\noindent\textfonts\rm Submitted to \journal\par 
     \bigskip\fi} 
\def\today{\number\day\ \ifcase\month\or 
     January\or February\or March\or April\or May\or June\or 
     July\or August\or September\or October\or November\or 
     December\fi\space \number\year} 
\def\date{\ifppt\noindent\textfonts\rm  
     Date: \today\par\goodbreak\bigskip\fi} 
% 
% Physics Abstracts classification numbers 
% 
\def\pacs#1{\ifppt\noindent\textfonts\rm  
     PACS number(s): #1\par\bigskip\fi} 
% 
 
% 
%%%%%%%%%%%%%%%%%%%%%%%%%%%%%%%%%%%%%%%%%%%%%%%%%%%%%%%%%%%% 
%                                                          % 
%  Sections, subsections, etc                              % 
%                                                          % 
%%%%%%%%%%%%%%%%%%%%%%%%%%%%%%%%%%%%%%%%%%%%%%%%%%%%%%%%%%%% 
% 
\def\section#1{\ifppt\ifnum\secno=0\eject\fi\fi 
    \subno=0\subsubno=0\global\advance\secno by 1 
    \gdef\labeltype{\seclabel}\ifnumbysec\countno=1\fi 
    \goodbreak\beforesecspace\nobreak 
    \noindent{\bf \the\secno. #1}\par\futurelet\next\sp@ce} 
\def\subsection#1{\subsubno=0\global\advance\subno by 1 
     \gdef\labeltype{\seclabel}% 
     \ifssf\else\goodbreak\beforesubspace\fi 
     \global\ssffalse\nobreak 
     \noindent{\it \the\secno.\the\subno. #1\par}% 
     \futurelet\next\subsp@ce} 
\def\subsubsection#1{\global\advance\subsubno by 1 
     \gdef\labeltype{\seclabel}% 
     \ifssf\else\goodbreak\beforesubsubspace\fi 
     \global\ssffalse\nobreak 
     \noindent{\it \the\secno.\the\subno.\the\subsubno. #1}\null.  
     \ignorespaces} 
% 
 
% 
%%%%%%%%%%%%%%%%%%%%%%%%%%%%%%%%%%%%%%%%%%%%%%%%%%%%%%%%%%%% 
%                                                          % 
%  Appendices                                              % 
%                                                          % 
%%%%%%%%%%%%%%%%%%%%%%%%%%%%%%%%%%%%%%%%%%%%%%%%%%%%%%%%%%%% 
% 
\def\numappendix#1{\ifappendix\ifnumbysec\countno=1\fi\else 
    \countno=1\figno=0\tabno=0\fi 
    \subno=0\global\advance\appno by 1 
    \secno=\appno\gdef\applett{A}\gdef\labeltype{\seclabel}% 
    \global\appendixtrue\global\numapptrue 
    \goodbreak\beforesecspace\nobreak 
    \noindent{\bf Appendix \the\appno. #1\par}% 
    \futurelet\next\sp@ce} 
\def\numsubappendix#1{\global\advance\subno by 1\subsubno=0 
    \gdef\labeltype{\seclabel}% 
    \ifssf\else\goodbreak\beforesubspace\fi 
    \global\ssffalse\nobreak 
    \noindent{\it A\the\appno.\the\subno. #1\par}% 
    \futurelet\next\subsp@ce} 
\def\@ppendix#1#2#3{\countno=1\subno=0\subsubno=0\secno=0\figno=0\tabno=0 
    \gdef\applett{#1}\gdef\labeltype{\seclabel}\global\appendixtrue 
    \goodbreak\beforesecspace\nobreak 
    \noindent{\bf Appendix#2#3\par}\futurelet\next\sp@ce} 
\def\Appendix#1{\@ppendix{A}{. }{#1}} 
\def\appendix#1#2{\@ppendix{#1}{ #1. }{#2}} 
\def\App#1{\@ppendix{A}{ }{#1}} 
\def\app{\@ppendix{A}{}{}} 
\def\subappendix#1#2{\global\advance\subno by 1\subsubno=0 
    \gdef\labeltype{\seclabel}% 
    \ifssf\else\goodbreak\beforesubspace\fi 
    \global\ssffalse\nobreak 
    \noindent{\it #1\the\subno. #2\par}% 
    \nobreak\subspace\noindent\ignorespaces} 
% 
%%%%%%%%%%%%%%%%%%%%%%%%%%%%%%%%%%%%%%%%%%%%%%%%%%%%%%%%%%%% 
%                                                          % 
%  Acknowledgments, notes added and foreign abstracts      % 
%                                                          % 
%%%%%%%%%%%%%%%%%%%%%%%%%%%%%%%%%%%%%%%%%%%%%%%%%%%%%%%%%%%% 
% 
\def\@ck#1{\ifletter\bigskip\noindent\ignorespaces\else 
    \goodbreak\beforesecspace\nobreak 
    \noindent{\bf Acknowledgment#1\par}% 
    \nobreak\secspace\noindent\ignorespaces\fi} 
\def\ack{\@ck{s}} 
\def\ackn{\@ck{}} 
\def\n@ip#1{\goodbreak\beforesecspace\nobreak 
    \noindent\smallfonts{\it #1}. \rm\ignorespaces} 
\def\naip{\n@ip{Note added in proof}} 
\def\na{\n@ip{Note added}} 
 
% 
%  \resume and \zus in Physics in Medicine and Biology only 
% 
 
% 
 
% 
%%%%%%%%%%%%%%%%%%%%%%%%%%%%%%%%%%%%%%%%%%%%%%%%%%%%%%%%%%%% 
%                                                          % 
%  Tables                                                  % 
%                                                          % 
%%%%%%%%%%%%%%%%%%%%%%%%%%%%%%%%%%%%%%%%%%%%%%%%%%%%%%%%%%% 
% 
 
% 
 
% 
\def\table#1{\tablecaption{#1}} 
\def\tablecont{\topinsert\global\advance\tabno by -1 
    \tablecaption{(continued)}} 
\def\tablecaption#1{\gdef\labeltype{\tablabel}\global\widefalse 
    \leftskip=\secindent\parindent=0pt 
    \global\advance\tabno by 1 
    \smallfonts{\bf Table \ifappendix\applett\fi\the\tabno.} \rm #1\par 
    \smallskip\futurelet\next\t@b} 
\def\endtable{\vfill\goodbreak} 
\def\t@b{\ifx\next*\let\next=\widet@b 
             \else\ifx\next[\let\next=\fullwidet@b 
                      \else\let\next=\narrowt@b\fi\fi 
             \next} 
\def\widet@b#1{\global\widetrue\global\notfulltrue 
    \t@bwidth=\hsize\advance\t@bwidth by -\secindent}  
\def\fullwidet@b[#1]{\global\widetrue\global\notfullfalse 
    \leftskip=0pt\t@bwidth=\hsize}                   
\def\narrowt@b{\global\notfulltrue} 
\def\align{\catcode`?=13\ifnotfull\moveright\secindent\fi 
    \vbox\bgroup\halign\ifwide to \t@bwidth\fi 
    \bgroup\strut\tabskip=1.2pc plus1pc minus.5pc} 
\def\endalign{\egroup\egroup\catcode`?=12} 
 
% 
% Use \L{#}, \R{#} and \C{#} to specify left, right or centred 
% columns immediately after \table. For example 
% \align\L{#}&&\L{#}\cr gives the preamble for a table with 
% all columns aligned left, \align\L{#}&\C{#}&\R{#}\cr 
% gives a table with 3 columns, the first aligned left, the second 
% centred and the third aligned right. 
% 
\def\L#1{#1\hfill}

% 
%  Rules for tables \br at top and bottom 
%  \mr to separate headings from entries 
% 
\def\br{\noalign{\vskip2pt\hrule height1pt\vskip2pt}} 
 
% 
 
% 
% Definitions for centring headings over several columns 
% \centre{4}{Results for helium} will centre 
% Results for helium over four columns 
% \crule{4} will produce a rule centred over four columns 
% to go below a centred heading 
% 
\def\centre#1#2{\multispan{#1}{\hfill#2\hfill}} 
\def\crule#1{\multispan{#1}{\hrulefill}} 

\catcode`?=13 
\def\lineup{\setbox0=\hbox{\smallfonts\rm 0}% 
    \digitwidth=\wd0% 
    \def?{\kern\digitwidth}% 
    \def\\{\hbox{$\phantom{-}$}}% 
    \def\-{\llap{$-$}}} 
\catcode`?=12 
% 
% Macros for two parts of a table of equal width side by side 
% \table{caption}[w] 
% \sidetable{first part}{second part} 
% \endtable 
% Use \table preamble for tables of 31picas width 
% 
\def\sidetable#1#2{\hbox{\ifppt\hsize=18pc\t@bwidth=18pc 
                          \else\hsize=15pc\t@bwidth=15pc\fi 
    \parindent=0pt\vtop{\null #1\par}% 
    \ifppt\hskip1.2pc\else\hskip1pc\fi 
    \vtop{\null #2\par}}}  
\def\lstable#1#2{\everypar{}\tempval=\hsize\hsize=\vsize 
    \vsize=\tempval\hoffset=-3pc 
    \global\tabno=#1\gdef\labeltype{\tablabel}% 
    \noindent\smallfonts{\bf Table \ifappendix\applett\fi 
    \the\tabno.} \rm #2\par 
    \smallskip\futurelet\next\t@b} 
\def\inctabno{\global\advance\tabno by 1} 
% 
%%%%%%%%%%%%%%%%%%%%%%%%%%%%%%%%%%%%%%%%%%%%%%%%%%%%%%%%%%%% 
%                                                          % 
%  Figures                                                 % 
%                                                          % 
%%%%%%%%%%%%%%%%%%%%%%%%%%%%%%%%%%%%%%%%%%%%%%%%%%%%%%%%%%%% 
% 
 
% 
 
% 
\def\figure#1{\figc@ption{#1}\bigskip} 
\def\figc@ption#1{\global\advance\figno by 1\gdef\labeltype{\figlabel}% 
   {\parindent=\secindent\smallfonts\hang 
    {\bf Figure \ifappendix\applett\fi\the\figno.} \rm #1\par}} 
% 
%%%%%%%%%%%%%%%%%%%%%%%%%%%%%%%%%%%%%%%%%%%%%%%%%%%%%%%%%%%% 
%                                                          % 
%  Reference lists                                         % 
%                                                          % 
%%%%%%%%%%%%%%%%%%%%%%%%%%%%%%%%%%%%%%%%%%%%%%%%%%%%%%%%%%%% 
% 
\def\refHEAD{\goodbreak\beforesecspace 
     \noindent\textfonts{\bf References}\par 
     \let\ref=\rf 
     \nobreak\smallfonts\rm} 
\def\numreferences{\refHEAD\parindent=30pt 
     \everypar{\hang\noindent\frenchspacing\rm} 
     \secspace} 
\def\rf#1{\par\noindent\hbox to 21pt{\hss #1\quad}\ignorespaces} 
% 
 
% 
 
% 
% reference to a journal article in numerical system 
% 
 
% 
% reference to a book or report in numerical system 
% 
 
% 
%%%%%%%%%%%%%%%%%%%%%%%%%%%%%%%%%%%%%%%%%%%%%%%%%%%%%%%%%%%% 
%                                                          % 
%  Theorems, lemmas, etc                                   % 
%                                                          % 
%%%%%%%%%%%%%%%%%%%%%%%%%%%%%%%%%%%%%%%%%%%%%%%%%%%%%%%%%%%% 
% 

% 
% NB \note#1 is used to give a Note (as opposed to a paper or letter) 
% in PMB therefore use commands \notes#1 for numbered Note 
% instead of \note  
% 

% 
\catcode`\@=12 
% 
% Parameter values for `Preprint' style  
% 
 
% 
% Parameter values for `Journal' style  
% 
\def\jnlstyle{\pptfalse\headsize{14}{18}% 
\textsize{10}{12}% 
\smallsize{8}{10} 
\textind=16pt} 
% 
% Parameter values for `Eleven point' style  
% 
 
% 
% Parameter values for `Large size' style  
% 
 
% 
\parindent=\textind 
%%%%%%%%%%%%%%%%%%%%%%%%%%%%%%%%%%%%%%%%%%%%%%%%%%%%%%%%%%%%%%%%%%%%%%%%%%%%%%%
\input xref
\input epsf
\eqnobysec
%
%\def\ps{\noalign{\vskip3pt}}
% positive space in tables, matrices 
% \ns for negative space
\def\received#1{\insertspace 
     \parindent=\secindent\ifppt\textfonts\else\smallfonts\fi 
     \hang{Received #1}\rm } 
%\def\appendix{\goodbreak\beforesecspace 
%     \noindent\textfonts{\bf Appendix}\secspace} 
%\headline={\ifodd\pageno{\ifnum\pageno=\firstpage\titlehead
%   \else\rrhead\fi}\else\lrhead\fi} 
%\def\lpm#1#2{LPM-#1-LT#2}
%\def\figure#1{\global\advance\figno by 1\gdef\labeltype{\figlabel}% 
%   {\parindent=\secindent\smallfonts\hang 
%    {\bf Figure \ifappendix\applett\fi\the\figno.} \rm #1\par}} 
%\def\endtable{\parindent=\textind\textfonts\rm\bigskip} 
%\def\rrhead{\textfonts\hskip\secindent\it 
%    \shorttitle\hfill\folio} 
%\def\lrhead{\textfonts\hbox to\secindent{\folio\hss}% 
%   \it\aunames\hss} 
%\footline={\ifnum\pageno=\firstpage
%\smallfonts cond-mat/\hfil\textfonts\folio\fi}   
%\def\titlehead{\smallfonts 
%\hfil\lpm{}{}} 

%\firstpage=1
%\pageno=1

%\pptstyle
\jnlstyle

\jl{1}
    
\overfullrule=0pt

\title{Percolation and conduction in restricted geometries}[Percolation
and conduction in restricted geometries]

\author{P Lajk\'o\dag\ddag\ and L Turban\dag}[P Lajk\'o and L Turban]

\address{\dag Laboratoire de Physique des 
Mat\'eriaux, Universit\'e Henri Poincar\'e (Nancy~I),
BP~239,  F--54506~Vand\oe uvre l\`es Nancy Cedex, France}

\address{\ddag Institute of Theoretical Physics, Szeged University, 
H--6720 Szeged, Hungary}

\received{11 October 1999}

\abs
The finite-size scaling behaviour for percolation and conduction is studied
in two-dimensional triangular-shaped random resistor networks at the
percolation threshold. The numerical simulations are
performed using an efficient star-triangle algorithm. The percolation
exponents, linked to the critical behaviour at corners, are in
good agreement with the conformal results. The conductivity exponent,
$t'=\zeta/\nu$, is found to be independent of the shape of the system. Its
value is very close to recent estimates for the surface and
bulk conductivity exponents.  
\endabs 

%\vglue1cm

\pacs{05.70.Jk, 64.60.Ak}

\submitted

\date
\section{Introduction}
Since the paper of Cardy \cite{cardy83} we know that, at a second-order
phase transition, the local critical behaviour can be influenced by the
shape of the system. Furthermore, in the case of conformally
invariant two-dimensional (2D) systems, the tools of conformal invariance
can be used, at the critical point, to relate the local critical behaviour
at a corner to the surface critical behaviour \cite{cardy84,barber84}. The
corner shape, which is  scale invariant, leads to local exponents varying
continuously with the opening angle. This marginal local critical behaviour
has been indeed observed numerically for different systems
(see~\cite{igloi93} for a review) and, more recently, analytical results
have been obtained for the Ising model [5--8]. 

In this paper, we present a numerical study
of the critical behaviour of 2D random resistor networks with triangular
shapes. This problem involves two sets of exponents, namely the percolation
exponents and the conductivity exponents [9--11]. Although our main
interest concerns the conduction exponents, our simulations
allow us, as a by-product, to check the conformal predictions for the
corner exponents of the percolation problem.

The percolation exponents are known exactly, both in the bulk and at the
surface, through a correspondance with the limit $q\to1$ of the $q$-state
Potts model [12--14]. The conformal aspects of the critical percolation
problem in finite geometries have been extensively studied in
\cite{langlands94}, following the work of~\cite{cardy92}.
Conformal invariance has been also verified in a transfer-matrix
calculation of the surface percolation exponent, using the gap-exponent
relation \cite{dequeiroz95}. The critical behaviour at surface and corners
has been considered in~\cite{wolf90} where the conduction problem
is addressed briefly.

A recent series expansion study \cite{essam96} suggests that
the surface conductivity has the same critical behaviour as the
bulk one (see \cite{grassberger99} and references therein). The main purpose
of this work is to examine, with high numerical accuracy, wether the shape
of a finite system may have some influence on the scaling behaviour of the
conductivity.

We study the finite-size-scaling behaviour of the conductance
and percolation probability between points located at the corners of a
triangle, either on the triangular or on the square lattice. As in the
Lobb-Frank algorithm \cite{lobb84}, the numerical technique involves a
succession of triangle-star and star-triangle transformations, which allows
one to reduce the triangular resistor network to a star network in a finite
number of steps. 

In section~2 we introduce the different correlation functions of the
percolation and conduction problems and define the associated exponents. We
also review the conformal results for the corner exponents. In section~3 we
give a detailed explanation the triangle-star star-triangle algorithm. The
finite-size scaling simulation results are presented in section 4 and
discussed in section 5.

\section{Percolation and conduction correlation functions}
We consider a random resistor network for which each lattice bond has
a probability $p$ to have a unit conductance and $1-p$ to be an
insulator. Let the connectedness characteristic function $c_{ij}$ be
defined as 
$$
c_{ij}=\cases{$1$ & if sites $i$ and $j$ are connected\cr
$0$ & otherwise.}
\label{e2.1}
$$
With $[\cdots]_{\rm av}$ denoting a configurational average, the
percolation correlation function  
$$
P_{ij}=[c_{ij}]_{\rm av}
\label{e2.2}
$$
gives the probability that sites $i$ and $j$ belong to the same cluster of
conducting bonds. The average conductance given by
$$
G_{ij}=[g_{ij}]_{\rm av}
\label{e2.3}
$$
where $g_{ij}$ is the conductance of the system between sites $i$ and
$j$, plays the role of a correlation function, or non-local conductive
susceptibility, for the conduction problem \cite{fish78}. Let $N$ be
the number of samples taken into account in the configurational average and
$N_{\rm con}$ the corresponding number of samples for which sites $i$ and
$j$ are connected. The correlation functions in equations~\ref{e2.2}
and~\ref{e2.3} can be rewritten explicitly as   
$$
\eqalign{
&P_{ij}=\lim_{N\to\infty}{1\over N}\sum_{\alpha=1}^N c_{ij}^\alpha
=\lim_{N\to\infty}{N_{\rm con}\over N}\cr
&G_{ij}=\lim_{N\to\infty}{1\over N}\sum_{\alpha=1}^N 
g_{ij}^\alpha=\lim_{N\to\infty}{N_{\rm con}\over
N}{1\over N_{\rm con}}\sum_{\alpha=1}^{N_{\rm con}}[g_{ij}^{\rm
con}]^\alpha\,.\cr } \label{e2.4}
$$
Thus one can define the reduced conduction correlation function
$$
\Gamma_{ij}={G_{ij}\over P_{ij}}=\lim_{N\to\infty}{1\over N_{\rm
con}} \sum_{\alpha=1}^{N_{\rm con}}[g_{ij}^{\rm con}]^\alpha
\label{e2.5}
$$
which gives the average conductance between two points $i$ and $j$
when they belong to the same percolation cluster.

In an infinite system, for two points at a distance $r_{ij}=r$, the
correlation functions display a power law decay  
$$
P_{ij}\sim r^{-2x}\qquad \Gamma_{ij}\sim r^{-\zeta/\nu}
\label{e2.6}
$$
at the percolation threshold $p=p_{\rm c}$. The exponents $x$ and $\nu$ are
the scaling dimension of the bulk order parameter and the correlation
length exponent for the percolation problem, respectively. The
conductivity exponent $\zeta$ governs the behaviour of the
macrosopic conductivity near the percolation threshold
where \cite{degennes76} 
$$
\Sigma\sim (p-p_{\rm c})^t\qquad t=\nu t'=\zeta+(d-2)\nu\qquad p>p_{\rm
c}\,. 
\label{e2.7}
$$
Note that $\zeta/\nu=t'$ in two dimensions.

As mentioned in the introduction, the percolation exponents are exactly
known in two dimensionsthrough a correspondance with the $q$-state Potts
model \cite{wu82,cardy87}: 
$$ 
\nu=\case{4}{3}\qquad x=\case{5}{48}\qquad
x_{\rm s}=\case{1}{3} 
\label{e2.8}
$$
where $x_{\rm s}$ is the scaling dimension of the surface order parameter
at the ordinary transition. 

Recent high statistics simulations led to the following accurate
estimate for the conductivity exponent in two dimensions
\cite{grassberger99}:  
$$
t'={\zeta\over\nu}=0.9825\pm0.0008\qquad \zeta=1.3100\pm0.0011\,.
\label{e2.9}
$$
Low-density series expansion results \cite{essam96} are
consistent with a surface conductivity exponent $\zeta_{\rm s}$
keeping its bulk value $\zeta$.  

For a triangular-shaped system of size $L$,
when the points $i$ and $j$ are located at corners with opening angles
$\theta_i$ and $\theta_j$, according to finite-size scaling, one expects the
following behaviour at criticality: 
$$
\eqalign{
&P_{ij}=P(\theta_i,\theta_j;L)\sim
L^{-\eta(\theta_i,\theta_j)}\sim L^{-x(\theta_i)-x(\theta_j)}\cr
&\Gamma_{ij}=\Gamma(\theta_i,\theta_j;L) \sim
L^{-\zeta(\theta_i,\theta_j)/\nu}\,.\cr} 
\label{e2.10}
$$
Here $x(\theta_i)$ is the scaling dimension of the local order parameter at a
corner with opening angle $\theta_i$. In the following, we shall also consider
the three-point correlation function $P_{ijk}$, which gives the probability that
the points $i$, $j$ and $k$, located at corners with opening angles $\theta_i$,
$\theta_j$ and $\theta_k$, belong to the same cluster. This quantity scales like
the product of the corresponding local order parameters, i.e., as 
$$
P_{ijk}=P(\theta_i,\theta_j,\theta_k;L)\sim
L^{-\eta(\theta_i,\theta_j,\theta_k)}\sim
L^{-x(\theta_i)-x(\theta_j)-x(\theta_k)}
\label{e2.11} 
$$
at the percolation threshold.

%%%%%%%%%%%%%%%%%%%%%%%%%%% figure 1
{\par\begingroup\medskip
\epsfxsize=11truecm
\topinsert
\centerline{\epsfbox{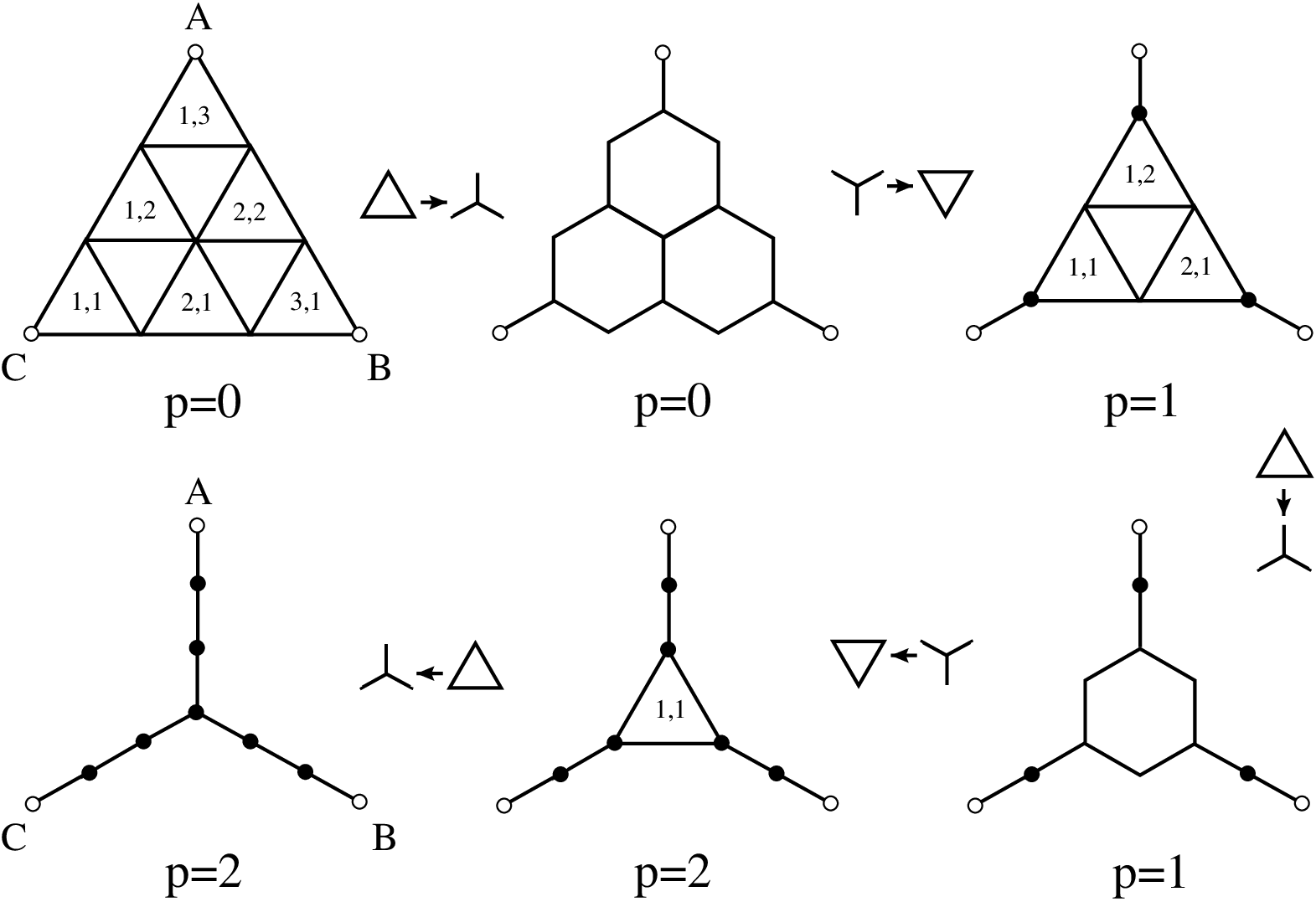}}
\smallskip
\figure{Reduction of a triangular-shaped system to a star
through a succession of triangle-star and star-triangle
transformations.\label{fig1}} 
\endinsert  
\endgroup
\par}

%%%%%%%%%%%%%%%%%%%%%%%%%%%%%%% figures 2 and 3
{\par\begingroup\medskip
\epsfxsize=6.5truecm
\topinsert
\centerline{\epsfbox{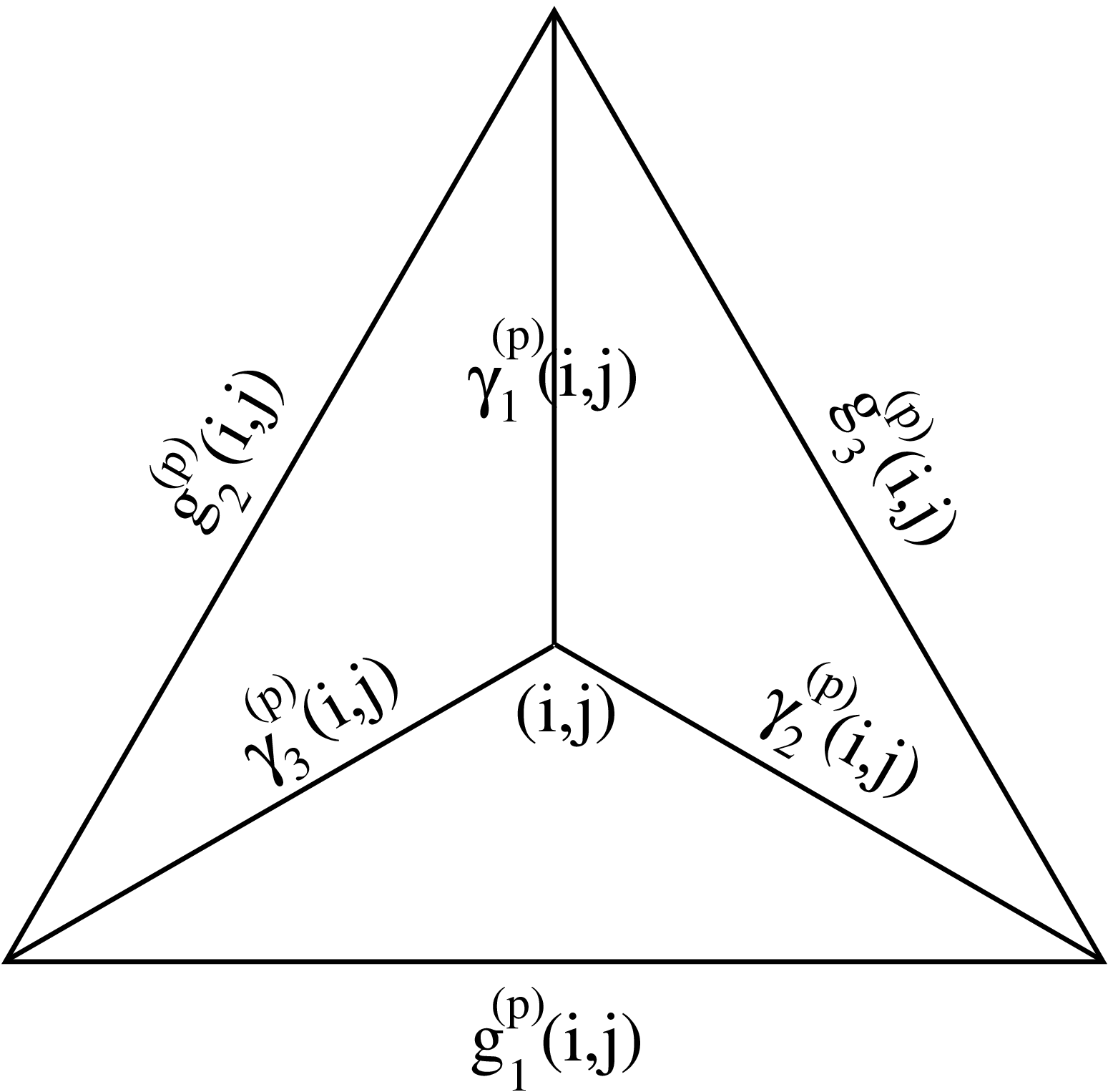}}
\smallskip
\figure{In the triangle-star transformation, up-pointing triangles
with con\-duc\-tances $g_\alpha^{(p)}(i,j)$ are transformed into
stars with conductances $\gamma_\alpha^{(p)}(i,j)$ given by
equation~(3.1).\label{fig2}}  
\epsfxsize=9.5truecm
\centerline{\epsfbox{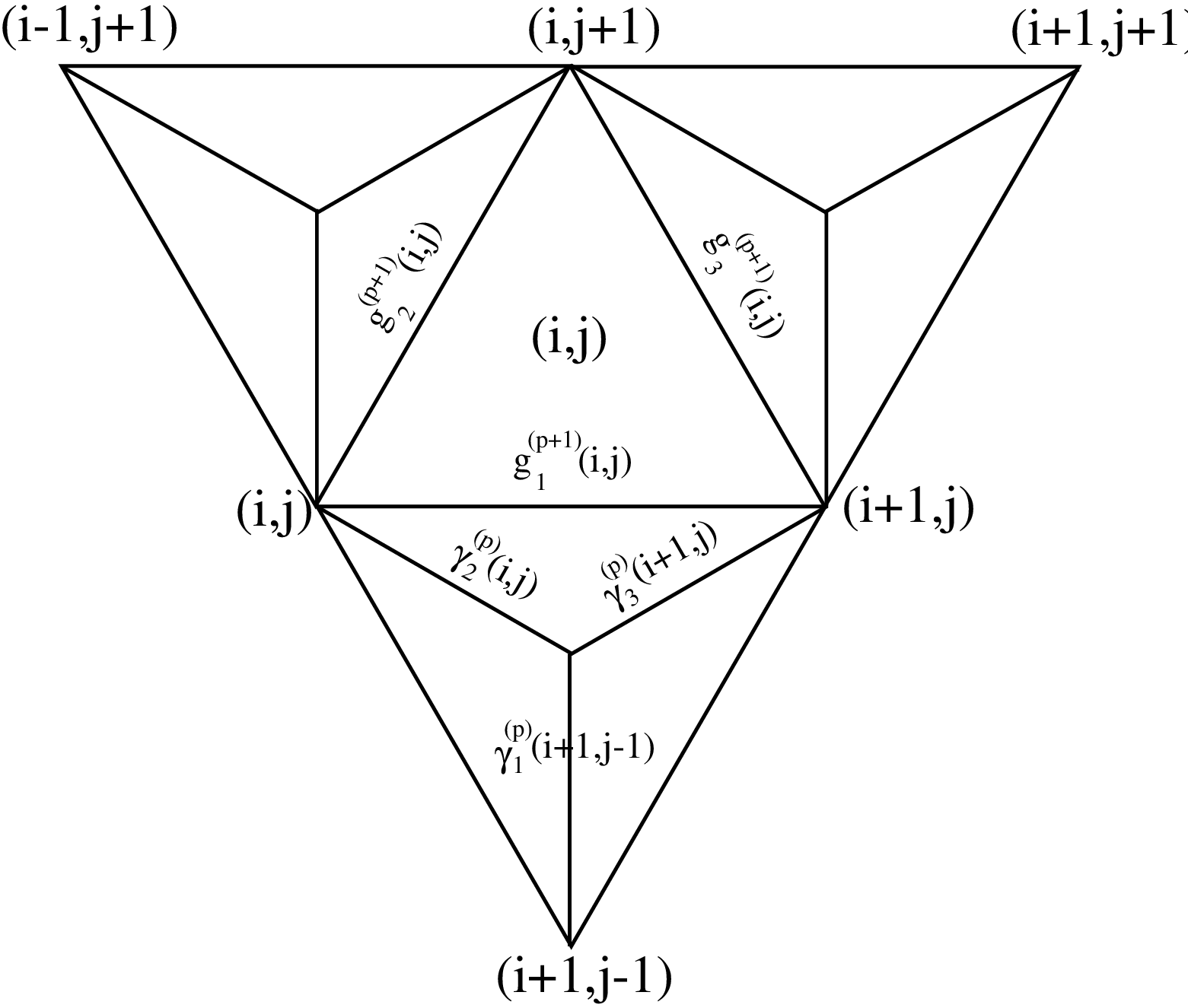}}
\smallskip 
\figure{The conductances $g_\alpha^{(p+1)}(i,j)$ on up-pointing triangles
at step $p+1$, given in equation~(3.2), follow from star-triangle
transformations on down-pointing stars. The conductances of the star involved in
the construction of $g_1^{(p+1)}(i,j)$ are indicated. They originate from
different up-pointing triangles at step $p$.\label{fig3}}   
\endinsert   
\endgroup 
\par}

A dependance of the local exponents on the opening angles is
generally expected since a wedge is a scale-invariant geometry and the angles
are marginal variables for the local critical behaviour. The critical 2D
Potts model being conformally invariant, one obtains the local critical
behaviour for the percolation problem at a corner using the conformal
transformation $w=z^{\theta/\pi}$, which maps the half-space onto a wedge
with opening angle $\theta$ \cite{cardy84,barber84}. This leads to the
following expression for the scaling dimension of the order parameter:
$$
x(\theta)={\pi\over\theta}\, x_{\rm s}\,.
\label{e2.12}
$$

%%%%%%%%%%%%%%%%%%%%%%%%%%%%%% figure 4
{\par\begingroup\medskip
\epsfxsize=11truecm
\topinsert
\centerline{\epsfbox{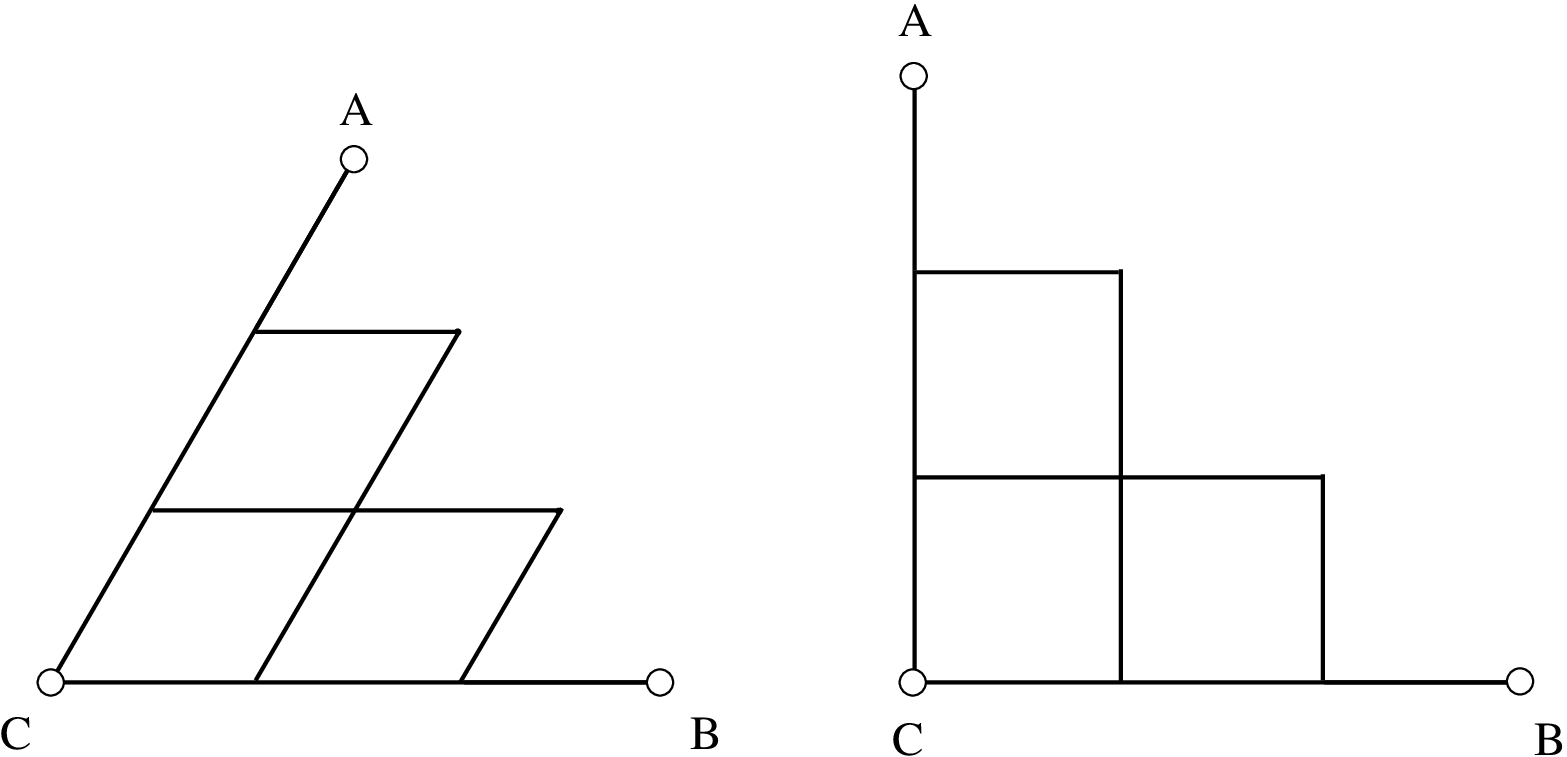}}
\smallskip
\figure{A triangular-shaped system on the square lattice is obtained
through a deformation of the triangular lattice when the
conductances are set to zero in one of the principal
directions.\label{fig4}}   
\endinsert    
\endgroup  
\par}

\section{The star-triangle transformation}
Let us consider a finite random resistor network with the shape of an
equilateral triangle of side $L$ on the triangular lattice. Through a succession
of triangle-star and star-triangle transformations, the original system can be
progressively transformed into a star with 3 branches of length $L$ as
shown in figure~\ref{fig1}.

At step $p$, in the first part of the lattice transformation, up-pointing
triangles with coordinates $(i,j)$ and bond conductances
$g_\alpha^{(p)}(i,j)$ ($\alpha=1,2,3$) are  replaced by stars with
bond conductances  
$$
\fl
\gamma_\alpha^{(p)}(i,j)\!=\!{g_1^{(p)}(i,j)\, 
g_2^{(p)}(i,j)\!+\! g_2^{(p)}(i,j)\, g_3^{(p)}(i,j)\!+\! g_3^{(p)}(i,j)\,
g_1^{(p)}(i,j)\over g_\alpha^{(p)}(i,j)}\quad\alpha=1,2,3
\label{e3.1}
$$
as shown in figure~\ref{fig2}. In the second part, down-pointing
stars are transformed into down-pointing triangles and
up-pointing triangles are relabelled as indicated in
figure~\ref{fig3}. Thus one obtains the bond conductances for
up-pointing triangles at step $p+1$ as
$$
\eqalign{
&g_1^{(p+1)}(i,j)={\gamma_2^{(p)}(i,j)\,\gamma_3^{(p)}(i+1,j)\over
\gamma_1^{(p)}(i+1,j-1)+\gamma_2^{(p)}(i,j)+\gamma_3^{(p)}(i+1,j)}\cr 
&g_2^{(p+1)}(i,j)={\gamma_3^{(p)}(i,j+1)\,\gamma_1^{(p)}(i,j)\over
\gamma_1^{(p)}(i,j)+\gamma_2^{(p)}(i-1,j+1)+\gamma_3^{(p)}(i,j+1)}\cr
&g_3^{(p+1)}(i,j)={\gamma_1^{(p)}(i+1,j)\,\gamma_2^{(p)}(i,j+1)\over
\gamma_1^{(p)}(i+1,j)+\gamma_2^{(p)}(i,j+1)+\gamma_3^{(p)}(i+1,j+1)}\,.\cr
}
\label{e3.2}
$$
Note that surface bonds in up-pointing triangles at step
$p+1$ result from the transformation of incomplete down-pointing stars.
The expressions given in~\ref{e3.2} still apply in this case, provided
the conductances associated with the missing bonds are set equal to
zero, i.e., with the boundary conditions
$$
\eqalign{
&\gamma_1^{(p)}(i,0)=0\qquad\gamma_2^{(p)}(0,j)=0\qquad
i,j=2,L-p\cr
&\gamma_3^{(p)}(i,j)=0\qquad i=2,L-p\qquad j=L-p-i+2\,.\cr
}
\label{e3.3}
$$

At step $p=L-1$, the final star configuration is obtained after the
triangle-star transformation has been performed. Then,  for example, the
conductance between points $A$ and $B$ on a sytem with size $L$ is
given by 
$$
g_{AB}=\left[\sum_{p=0}^{L-1}{1\over\gamma_1^{(p)}(1,L-p)}
+{1\over\gamma_2^{(p)}(L-p,1)}\right]^{-1}\,. \label{e3.4}
$$
The same transformation, with all the conductances $g_3^{(0)}(i,j)$
set to zero in the initial configuration, can be used to reduce to a star
a triangular-shaped system on the square lattice as shown in
figure~\ref{fig4}.

\section{Finite-size scaling results} 

We have studied the finite-size scaling behaviour of the
percolation and reduced conduction correlation functions between corners on
triangular-shaped sytems. We worked at the percolation threshold, either on the
triangular lattice ($p_{\rm c}=2\sin(\pi/18)$) or on the square
lattice ($p_{\rm c}=1/2$) \cite{shante71}. 

%%%%%%%%%%%%%%%%%%%%%%%%%%%%%%%%%%%% figures 5, 6 and 7
{\par\begingroup\medskip
\epsfxsize=7.5truecm
\topinsert
\centerline{\epsfbox{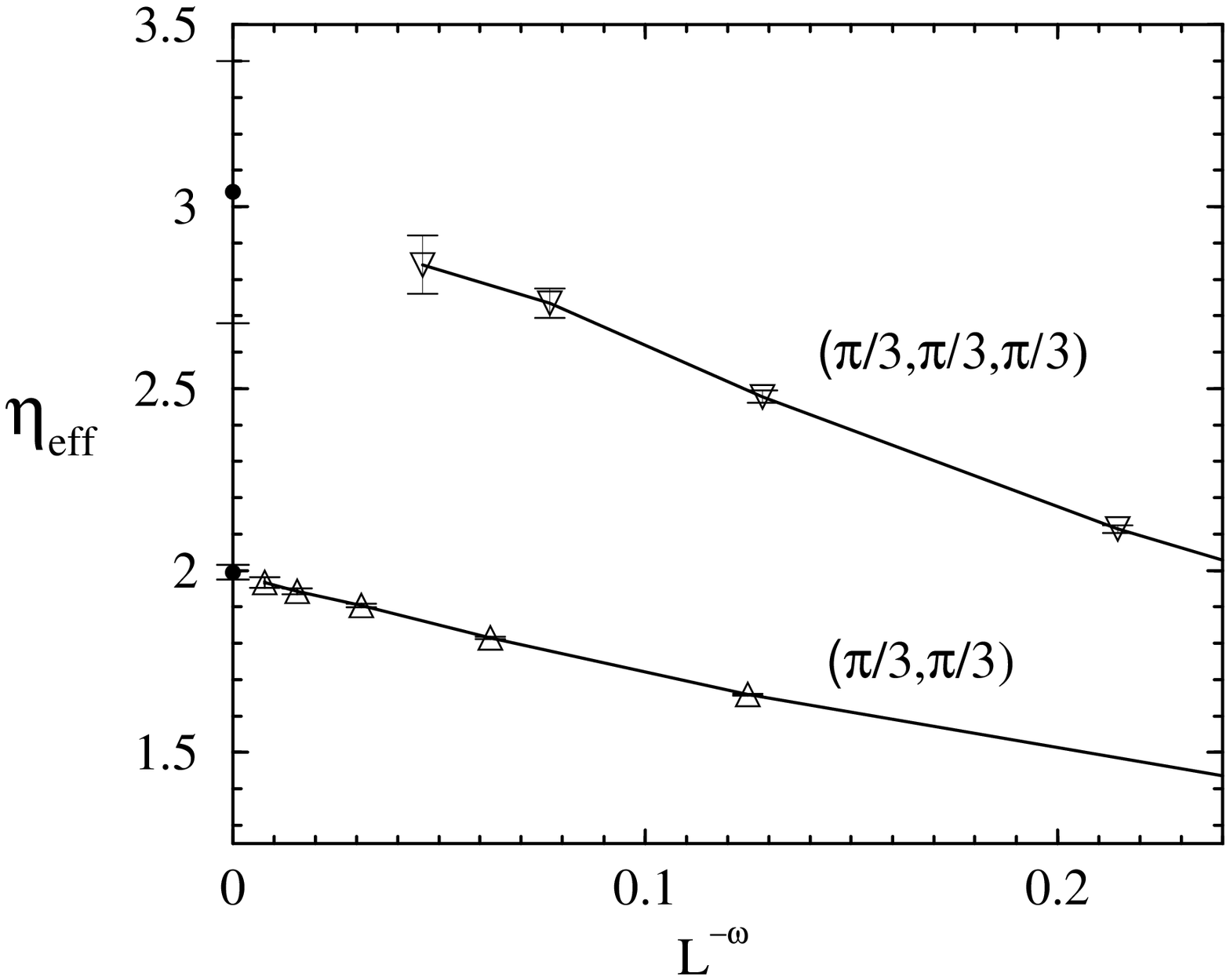}}
\smallskip
\figure{Percolation on the triangular
lattice: effective decay exponents for the two-point (\opentri)
and three-point (\opentridown) correlation functions plotted
against  $L^{-\omega}$ and their extrapolated values
(\fullcirc).\label{fig5}}     
\medskip
\epsfxsize=7.5truecm
\centerline{\epsfbox{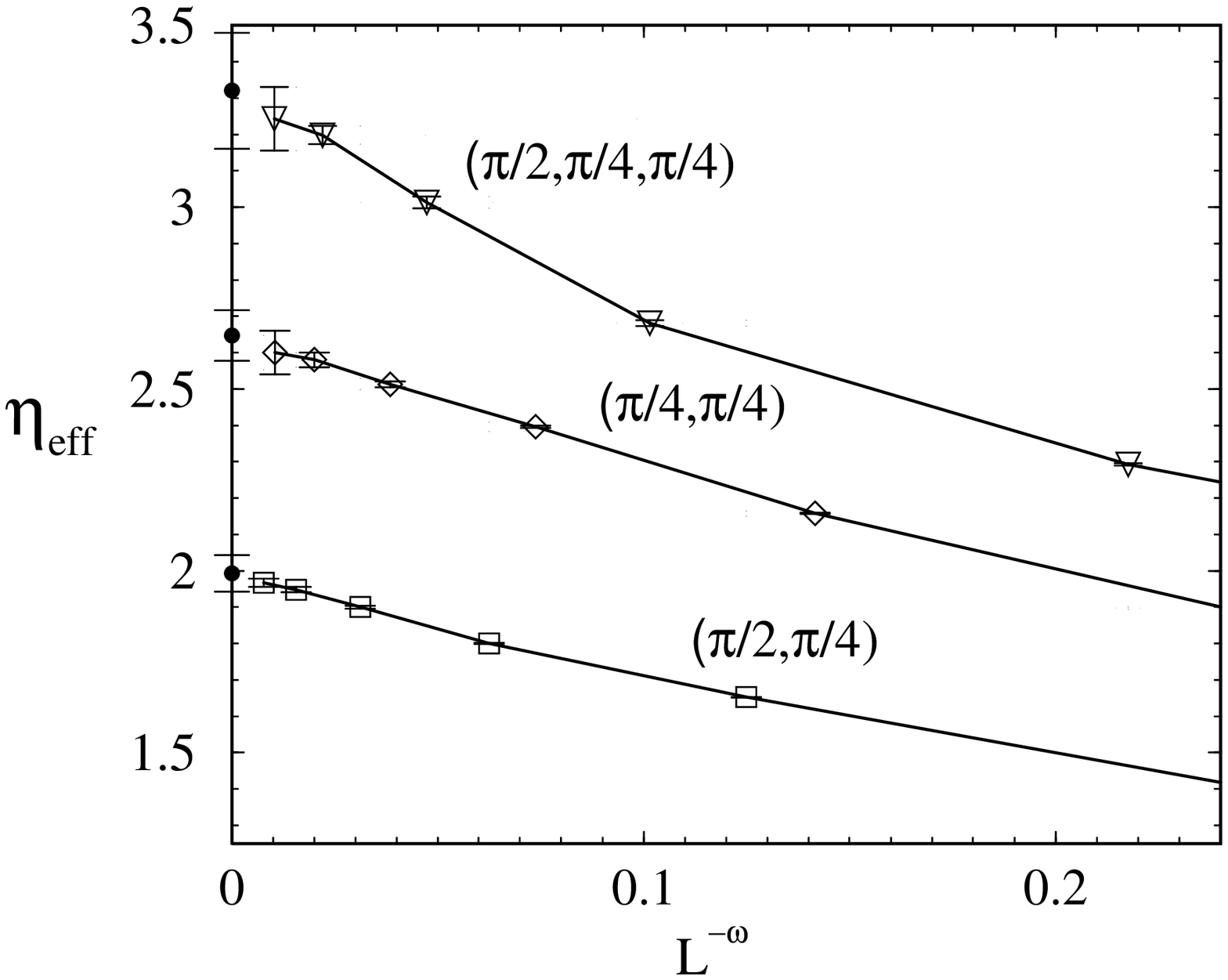}}
\smallskip
\figure{Percolation on the square
lattice: effective decay exponents for the two-point (\opensqr, \opendiamond)
and three-point (\opentridown) correlation functions plotted
against  $L^{-\omega}$ and their extrapolated values (\fullcirc).\label{fig6}}   
\endinsert    
\endgroup  
\par}

On the triangular lattice (figure~\ref{fig1}), we calculated the two-point
functions $P(\case{\pi}{3},\case{\pi}{3};L)$, $\Gamma(\case{\pi}{3},\case{\pi}{3};L)$ and the three-point
function $P(\case{\pi}{3},\case{\pi}{3},\case{\pi}{3};L)$ whereas on the square lattice
(figure \ref{fig4}), we studied the two-point functions $P(\case{\pi}{2},\case{\pi}{4};L)$,
$P(\case{\pi}{4},\case{\pi}{4};L)$, $\Gamma(\case{\pi}{2},\case{\pi}{4};L)$, $\Gamma(\case{\pi}{4},\case{\pi}{4};L)$ and the
three-point function $P(\case{\pi}{2},\case{\pi}{4},\case{\pi}{4};L)$. 

The initial bond configurations were generated using two types of random number
generators, a simple shift register algorithm and the ranlux97 generator. We
checked that both generators led to consistent results within the statistical
errors. On the square lattice, in order to increase the number of percolating
samples, the external bonds  connecting $A$ and $B$ to the rest of the
system in figure \ref{fig4} were always assumed to be conducting.
 
{\par\begingroup\medskip
\epsfxsize=7.5truecm
\topinsert
\centerline{\epsfbox{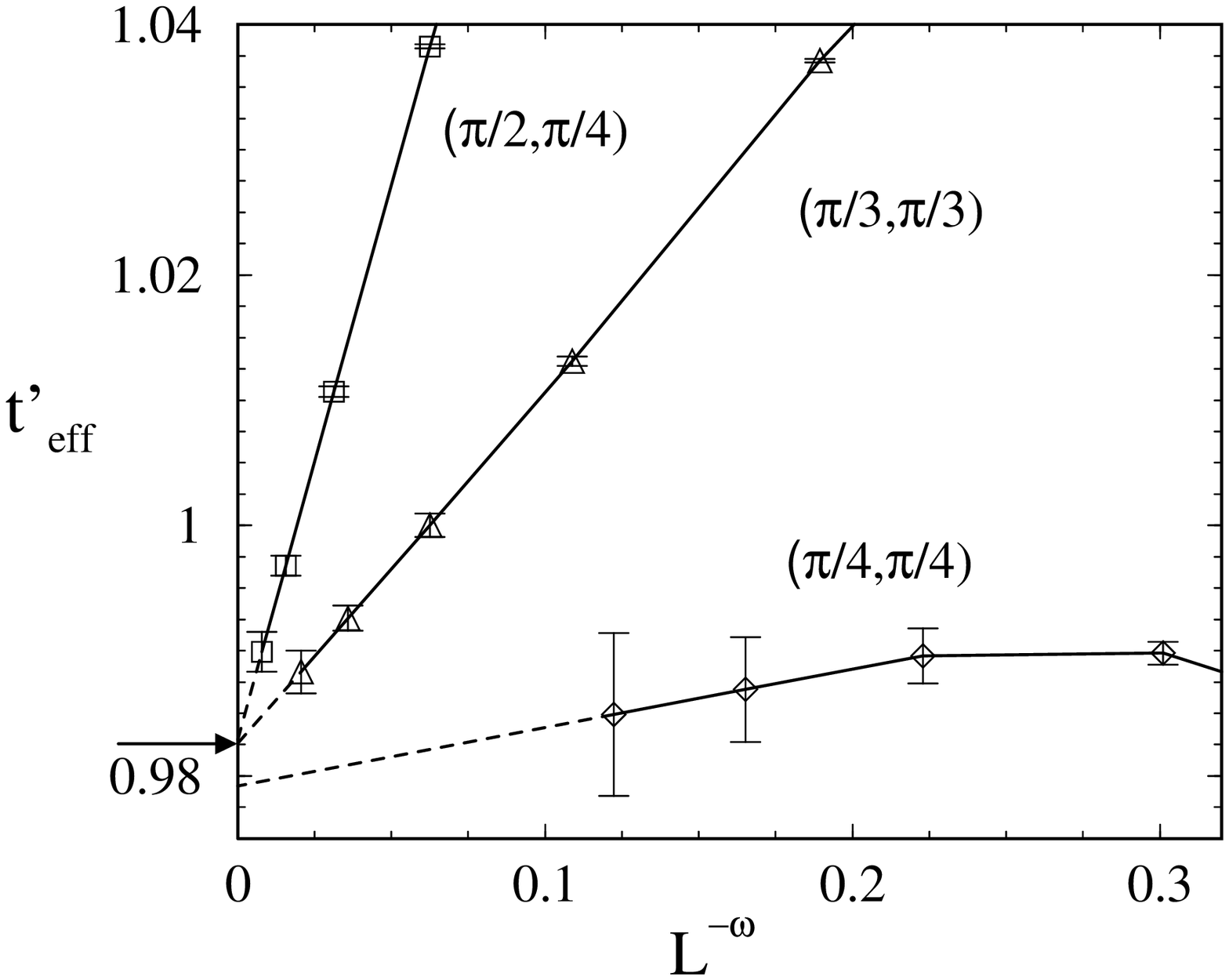}} \smallskip
\figure{Conduction in triangular-shaped
systems: effective decay exponents of the corner-to-corner
conduction correlation functions on the triangular (\opentri) and
the square (\opensqr, \opendiamond) lattices, plotted against $L^{-\omega}$.
The dotted lines are the best linear fits used for the
extrapolations and the arrow indicates Grassberger's
result for the the bulk conductivity exponent.\label{fig7}}      
\medskip
\table{Exponents governing the decay of the percolation correlation
functions ($\eta$) and the conduction correlation functions ($\zeta/\nu$)
for triangular-shaped systems at criticality on the triangular and
square lattices. The numerical values obtained for the percolation problem are
compared to the values expected from conformal invariance. The
last two lines give the effective correction-to-scaling
exponents $\omega$ used in the extrapolation process.\label{tab1}}[f]
\lineup 
\align\L{#}&&\L{#}\cr  
\br
&\centre{2}{Triangular lattice}&\centre{3}{Square lattice}\cr\ns
\crule{1}&\crule{2}&\crule{3}\cr
Angles& $(\case{\pi}{3},\case{\pi}{3})$ & $(\case{\pi}{3},\case{\pi}{3},\case{\pi}{3})$ & $(\case{\pi}{2},\case{\pi}{4})$ &
$(\case{\pi}{4},\case{\pi}{4})$ & $(\case{\pi}{2},\case{\pi}{4},\case{\pi}{4})$ \cr 
$\eta$ numerical & $1.99\pm0.02$ &$3.0\pm0.3$ & $2.01\pm0.05$ & $2.65\pm0.07$ &
$3.32\pm0.16$ \cr  
$\eta$ conf. inv. & $2$ & $3$ & $2$ & $\case{8}{3}$ & $\case{10}{3}$ \cr 
$\zeta/\nu$ numerical & $0.9827\pm0.0017$ & --- & $0.9829\pm0.0012$ &
$0.979\pm0.018$ & --- \cr 
$\omega$ percolation & $1.05$ & $0.76$ & $1.14$ & $1.06$ & $1.13$ \cr
$\omega$ conduction & $0.81$ & --- & $0.43$ &$1.02$ & --- \cr 
\br
\endalign
\endtable
\endinsert      
\endgroup   
\par}

We used system sizes of the form $L=2^k$ up to $L=256$. The star-triangle
algorithm described in section~3 led to a computation time scaling roughly
as $L^2 \ln L$. The number of samples generated was $N=2\times 10^7$ for the
triangular lattice, except for the largest size where $N=3\times 10^7$. On
the square lattice, $4\times 10^7$ samples were generated for all sizes. The
percolation and conduction correlation functions $P(L)$ and $\Gamma(L)$ were
stored as $n$ independent averages over groups of $10^6$ samples in order to
evaluate the statistical errors. More precisely, as defined in
equation~\ref{e2.5}, the conduction correlation function is averaged over
that part of the $10^6$ samples for which the two points are connected. 

Effective exponents at size $L$ for the percolation and conduction problems were
obtained using the two-points approximants
$$
\fl
\eta_{\rm eff}(L)={\ln P(L/2)-\ln P(2L)\over\ln4}\qquad
\zeta/\nu\vert_{\rm
eff}(L)={\ln\Gamma(L/2)-\ln\Gamma(2L)\over\ln4}\,. \label{e4.1}
$$
The central value of the approximants are calculated by averaging $P(L)$ and
$\Gamma(L)$ over all the samples. The error bars were deduced
from the mean-square  deviations $\sigma^2$ of the approximants, obtained
with the $n$ statistically independent averages, as
$\pm\sqrt{\sigma^2/(n-1)}$.

The finite-size results for the percolation exponents are shown in
figure~\ref{fig5} for the triangular lattice and figure~\ref{fig6} for the
square lattice. The results for the conductivity exponents are shown in
figure~\ref{fig7} for both lattices.  

The finite-size results were extrapolated in the following way. Assuming a single
correction-to-scaling exponent $\omega$, in each case we looked for the
value of $\omega$ leading to the best linear variation at large size for
the central value of the effective exponent as a function of $L^{-\omega}$.
The intercept with the vertical axis gives the extrapolated values
$\eta=\eta_{\rm eff}(\infty)$ and $\zeta/\nu=\zeta/\nu\vert_{\rm eff}(\infty)$. The
statistical error on the extrapolated values were deduced, as above, from the
mean-square deviation of the extrapolated exponents deduced from $n$
statistically independent sets of approximants, taking the same value of
$\omega$ for the linear fit. To the statistical error, we added a systematic
error, linked to the deviation from the asymptotic regime. It was taken as
the difference between the extrapolated values when one takes or not into
account the largest size in the extrapolation process. Doing so, we probably
overestimate the systematic error. 

Due to the non-monotonous behaviour of the effective exponent for the
conductance between corners with opening angle $\case{\pi}{4}$, the error
bar on the extrapolated exponent had to be estimated differently. It was
obtained through an extrapolation of the extreme values of the effective
exponents using the same method as for the central value. 

The extrapolated exponents for the percolation and conduction problems are shown
in table~\ref{tab1}.

\section{Discussion}

Let us first consider the percolation problem. The decay exponents of the
two- and three-point correlation functions, following from conformal
invariance, are easily obtained using equations \ref{e2.8} and
(2.10)--(2.12). Our numerical results in table \ref{tab1} are in good
agreement with the expected ones, although with a lower precision for the
three-point exponents, due to larger statistical errors. 

The percolation probability between two corners $P(\theta_i,\theta_j;L)$ can
be written as the sum of two contributions. The leading one is the
probability that $i$ and $j$ are connected without being connected to the
third corner $k$ which scales as $L^{-\eta(\theta_i,\theta_j)}$. The second
is the probability that the three corners are connected and it decays as
$L^{-\eta(\theta_i,\theta_j,\theta_k)}$. Thus we have a
correction-to-scaling exponent equal to $x(\theta_k)$ for the two-point
percolation probability. Another correction is due to the leading
irrelevant operator with scaling dimension $-1$ at the surface
\cite{dequeiroz95}. This explains the value, close to $1$, of the effective
correction exponent $\omega$ for two-point percolation, since the
amplitude of the first correction is small.   

Our main result concerns the decay exponent $\zeta/\nu$ which is
equal to $t'$ in two dimensions. The numerical values given in table
\ref{tab1} for the different geometries are all quite close to the value of
$t'$ for the bulk, which is given in equation \ref{e2.9}. These results,
together with the surface ones \cite{essam96}, strongly support the
existence of a single conductivity scale, independent of the sample
geometry. The influence of the opening angles can be seen in the amplitudes
and perhaps also in the correction-to-scaling exponents. 

One may notice that, apart from the case of the diagonal direction on the
square lattice, the approximants for the conductivity exponents display a
monotonous behaviour at large size. This is to be compared to the
non-monotonous behaviour observed in reference \cite{grassberger99} for bond
percolation conductivity on the square lattice. It makes the
extrapolation of the exponents more easy, thus reducing the computational
effort necessary for a given precision on the extrapolated values. 

Except for the diagonal direction on the square lattice, our error bars are
about the double of the ones of Grassberger, for a maximum system size
$L=256$ instead of $L=4096$ in \cite{grassberger99}. If one accepts the
universality of the conductivity exponent, which is strongly suggested by
our results, the triangular geometry used here appears as a
potentially efficient tool to further improve the precision on the value of
$\zeta/\nu$.

\ack  The Laboratoire de Physique des Mat\'eriaux is Unit\'e Mixte de
Recherche CNRS No 7556. The authors are indebted to F Igl\'oi for useful
discussions. This work has been supported by the Hungarian National
Research Fund under grants Nos OTKA F/7/026004, M028418 and by the
Hungarian Ministery of Education under grant No FKFP 0596/1999. PL
thanks the Soros Foundation, Budapest, for a travelling grant.

\numreferences
\bibitem{cardy83} {Cardy J L 1983}\ {\JPA}\ {\bf 16}\ {3617}
\bibitem{cardy84} {Cardy J L 1984}\ {\NP\it B}\ {\bf 240}\ {514}
\bibitem{barber84} {Barber M N, Peschel I and Pearce P A 1984}\ {\it J.
Stat. Phys.}\ {\bf 37}\ {497} 
\bibitem{igloi93} {Igl\'oi F, Peschel I and Turban L 1993}\ {\it Adv. 
Phys.}\ {\bf 42}\ {683}
\bibitem{abraham94} {Abraham D B and Latr\'emoli\`ere F T 1994}\ {\PR\
E}\ {\bf 50}\ {R9} 
\bibitem{abraham95} {Abraham D B and Latr\'emoli\`ere F T  1995}\ {\it J.
Stat. Phys.}\ {\bf 81}\ {539}  
\bibitem{abraham96} {Abraham D B and Latr\'emoli\`ere F T 1996}\ {\PRL}\
{\bf 76}\ {4813}  
\bibitem{davies97} {Davies B and Peschel I 1997}\ {\AP}\
{\bf 6}\ {187} 
\bibitem{kirkpatrick73} {Kirkpatrick S 1973}\ {\RMP}\ {\bf 45}\ {574}
\bibitem{bunde91} {Bunde A and Havlin S 1991}\ {\it Fractals and
Disordered Systems}\ {(Berlin: Springer)}\ {p 97} 
\bibitem{stauffer92} {Stauffer D and Aharony A 1992}\
{\it Introduction to Percolation Theory}\ {(London: Taylor \& Francis)}\ {p
89} 
\bibitem{kasteleyn69} {Kasteleyn P W and Fortuin C M 1969}\
{\it J. Phys. Soc. Japan. (Suppl.)}\ {\bf 26}\ {11}  
\bibitem{stephen77} {Stephen M J 1977}\ {\PR\ B}\ {\bf 15}\ {5674} 
\bibitem{wu78} {Wu F Y 1978}\ {\it J. Stat. Phys.}\ {\bf 18}\ {115} 
\bibitem{langlands94} {Langlands R, Pouliot P and Saint-Aubin Y 1994}\ {\it 
Bull. Am. Math. Soc.}\ {\bf 30}\ {1}
\bibitem{cardy92} {Cardy J L 1992}\ {\JPA}\ {\bf 25}\ {L201}
\bibitem{dequeiroz95} {de Queiroz S L A 1995}\ {\JPA}\ {\bf 27}\ {L363}
\bibitem{wolf90} {Wolf T, Blender R and Dietrich W 1990}\ {\JPA}\ {\bf 23}\
{L153} \bibitem{essam96} {Essam J W, Lookman T and De'Bell K 1996}\ {\JPA}\ 
{\bf 29}\ {L143}
\bibitem{grassberger99} {Grassberger P 1999}\ {\it Physica A}\   {\bf 262}\
{251} 
\bibitem{lobb84} {Lobb C J and Frank D J 1984}\ {\PR\ B}\ {\bf 30}\ {4090}
\bibitem{fish78} {Fish R and Harris A B 1978}\ {\PR\ B}\ {\bf 18}\ {416} 
\bibitem{degennes76} {de Gennnes P G 1976}\ {\it J. Physique Lett.}\ {\bf
37}\ {L1} 
\bibitem{wu82} {Wu F Y 1982}\ {\RMP}\ {\bf 54}\ {235}
\bibitem{cardy87} {Cardy J L 1987}\ {\it Phase transitions and critical
phenomena}\ {vol 11, ed C Domb and J L Lebowitz (New-York: Academic)}\ {p
55}
\bibitem{shante71} {Shante V K S and Kirkpatrick S 1971}\ {\it Adv. Phys}\
{\bf 20}\ {325}

\vfill\eject\bye